\documentclass[a4paper,fleqn]{cas-sc}
\AtBeginEnvironment{table}{\small}
\usepackage[numbers]{natbib}
\usepackage{graphicx} 
\usepackage{caption}
\usepackage{tikz}
\usetikzlibrary{positioning, shapes}
\usetikzlibrary{trees}
\usepackage{multirow}
\usepackage{pgfplots}
\pgfplotsset{compat=1.18}
\usepackage{longtable}
\usepackage{enumitem}
\usepackage{booktabs}
\usepackage{tabularx}
\usepackage{ragged2e}
\usepackage{pdflscape}
\usepackage{geometry}
\usepackage{enumitem}
\usepackage{array}
\usepackage{xcolor}

\usepackage{dirtytalk}

\usepackage[most]{tcolorbox}
\def\tsc#1{\csdef{#1}{\textsc{\lowercase{#1}}\xspace}}
\tsc{WGM}
\tsc{QE}
\tsc{EP}
\tsc{PMS}
\tsc{BEC}
\tsc{DE}

\begin{document}
\let\WriteBookmarks\relax
\def\floatpagepagefraction{1}
\def\textpagefraction{.001}

\shorttitle{CSE task recommendation}

\shortauthors{S. Nirmani et~al.}

\title [mode = title]{A Systematic Literature Review  on Task Recommendation Systems for Crowdsourced Software Engineering}                      

%
\author[1]{Shashiwadana Nirmani}[]

\cormark[1]


\ead{s.dona@research.deakin.edu.au}

\affiliation[1]{organization={Deakin University},
    addressline={Burwood}, 
    city={VIC},
    country={Australia}}

\author[2]{Mojtaba Shahin}[]
\ead{mojtaba.shahin@rmit.edu.au}

\author[1]{Hourieh Khalajzadeh}[]
\ead{hourieh.khalajzadeh@deakin.edu.au}

\affiliation[2]{organization={RMIT
University},
    addressline={Melbourne}, 
    city={VIC},
    country={Australia}}

\author%
[1]
{Xiao Liu}[style=chinese]
\ead{xiao.liu@deakin.edu.au}

\cortext[cor1]{Corresponding author}

\begin{abstract}
\noindent\textbf{Context}: Crowdsourced Software Engineering (CSE) offers outsourcing work to software practitioners by leveraging a global online workforce. However, these software practitioners struggle to identify suitable tasks due to the variety of options available. Hence, there have been a growing number of studies on introducing recommendation systems to recommend CSE tasks to software practitioners. 
\newline
\textbf{Objective}: The goal of this study is to analyze the existing CSE task recommendation systems, investigating their extracted data, recommendation methods, key advantages and limitations, recommended task types, the use of human factors in recommendations, popular platforms, and features used to make recommendations.
\newline
\textbf{Method}: This SLR was conducted according to the Kitchenham and Charters’ guidelines. We used manual and automatic search strategies without putting any time limitation for searching the relevant papers.
\newline
\textbf{Results}: We selected \textcolor{black} {65 primary studies} for data extraction, analysis, and synthesis based on our predefined inclusion and exclusion criteria. \textcolor{black}{Based on our data analysis results, we classified the extracted information into four categories according to the data acquisition sources: Software Practitioner’s Profile, Task or Project, Previous Contributions, and Direct Data Collection. We also organized the proposed recommendation systems into a taxonomy and identified key advantages, such as increased performance, accuracy, and optimized solutions. In addition, we identified the limitations of these systems, such as inadequate or biased recommendations and lack of generalizability.} Our results revealed that human factors play a major role in CSE task recommendation. Further, we identified five popular task types recommended, popular platforms, and their features used in task recommendation. We also provided recommendations for future research directions.
\newline
\textbf{Conclusion}: This SLR provides insights into current trends, gaps, and future research directions in CSE task recommendation systems \textcolor{black}{such as the need for comprehensive evaluation, standardized evaluation metrics, and benchmarking in future studies, transferring knowledge from other platforms to address cold start problem}. 
\end{abstract}

\begin{highlights}
\item Crowdsourced Software Engineering task recommendation is a trending research area
\item Content-based approaches dominate existing software task recommendation systems
\item Current recommendation systems in this field lack integration of human factors
\end{highlights}


\maketitle

\section{Introduction}
Crowdsourcing is a distributed problem-solving approach that combines human and machine computation. It involves organizations outsourcing work to an undefined networked labour pool through open calls for participation \cite{mao2017survey}. Crowdsourced Software Engineering (CSE) is derived from the crowdsourcing concept. CSE utilizes open calls to recruit global online labour in various software engineering tasks, including requirements extraction, design, coding, and testing \cite{mao2017survey}. It reduces time-to-market by enabling parallelism, lowering costs and defect rates, and enabling flexible development capabilities \cite{lakhani2010topcoder}\cite{latoza2013crowd}\cite{stol2014two}. Mao et al. state that there is a dramatic rise in recent work on the use of crowdsourcing in software engineering based on their conducted pilot study \cite{mao2017survey}.
\par
Popular crowdsourcing platform Topcoder \footnote{ \url{https://www.topcoder.com}} has hosted over 136,202 challenges in software design, development, and data science, with total prizes exceeding US\$191,873,091  \cite{topcoder_stats}. uTest \footnote{ \url{https://www.utest.com}}  has a freelancer pool of more than 1 million testers worldwide who test new apps for device compatibility, conduct functionality testing, and perform usability inspections \cite{utest_about}. As of April 2024, StackOverflow \footnote{ \url{stackoverflow.com}}  has over 24 million programming-related questions and 36 million answers  \cite{stack_exchange_data}. Major companies like Netflix, Microsoft, Facebook, and Google frequently offer bug bounties, particularly for security vulnerabilities \cite{latoza2015crowdsourcing}. With such extensive community involvement, it is crucial to recommend the most suitable tasks to the contributors.
\par
Developers often encounter various technical and social barriers when attempting to onboard crowdsourcing projects \cite{casalnuovo2015developer}. Unsuccessful onboarding not only affects developers but also impacts the projects themselves. It increases the time and effort new developers spend searching for suitable projects and learning \cite{matek2016github}. Further, online crowdsourced software development operates as a globally distributed collaborative effort. It involves a diverse range of contributors with varying personalities, educational backgrounds, and levels of expertise. These contributors work from different locations across multiple time zones and undertake various development tasks, such as bug fixing, code testing, or documentation, driven by their individual motivations. Hence, these contributors may struggle to find suitable projects or tasks among the thousands available in these communities \cite{yang2016repolike}. Various recommendation systems have been introduced in the literature to provide personalized task recommendations to software practitioners on crowdsourcing platforms. Providing such recommendations results in increased worker satisfaction, lower churning rate after onboarding, and improved task quality.
\par
\textcolor{black}{We retrieved relevant studies from 6 popular digital databases, including Scopus, IEEE Xplore, and ACM digital library, where we found 6, 20, 15, 5, 3, and 4 primary studies, respectively. After forward and backward snowballing, the total number of primary studies led to 65.  We found an existing SLR and a survey that are slightly relevant to the focused domain of this study. However, the SLR by Zhen et al. \cite{zhen2021crowdsourcing} focuses broadly on software as well as non-software tasks in crowdsourcing, while our SLR provides an in-depth analysis of personalized CSE task recommendations for software practitioners. Similarly, Mao et al. \cite{mao2017survey}'s survey broadly analyzes crowdsourced software engineering practices without specific research questions or overlapping studies, whereas our SLR focuses specifically on recommending tasks to practitioners. }

\par
\textcolor{black}{This SLR aims to consolidate existing information and provide an in-depth understanding of the existing literature on providing personalized task recommendations to software practitioners on crowdsourcing platforms. The findings of this study offer insights for both practitioners and researchers focused on enhancing the CSE task recommendation systems while also laying the groundwork for future research in this field.} The key contributions of this SLR are as follows:
\begin{itemize}
    \item Identifying, analyzing, extracting, and synthesizing \textcolor{black}{65 highly relevant} primary studies.
    \item Analysing how software task recommendation in crowdsourcing has been done so far and its current status.
    \item Identifying key advantages and limitations in the existing studies.
    \item Providing insights into key future research directions and recommendations for further studies.
    
\end{itemize}

\textcolor{black}{The key results of our study revealed that the data used to recommend tasks were sourced from 4
sources: software practitioner’s profile, task or project itself, previously contributed tasks or projects available
in the crowdsourcing platform and direct data collection via a customized platform. The existing software task recommendation methods can be mainly categorized as content-based, collaborative filtering, and hybrid approaches. Content-based approaches are the most popular category. 
The existing CSE task recommendation systems offer key advantages such as increased performance, accuracy, optimized solutions, and the ability to provide more comprehensive and diverse recommendations. However, they also have three main limitations: limitations in the dataset, approach, and evaluation. Commonly recommended tasks in CSE platforms, such as GitHub, Topcoder, and Utest, include repository contributions, issue fixes, testing tasks, Good First Issues (GFI), and answering questions. Among the various features used in existing systems, issue or task descriptions are the most widely utilized for data extraction. Key future research directions include the need for integrating more human factors into these systems, transferring knowledge from other platforms to overcome data sparsity, and exploring the possibility of integrating these recommendation systems with development workflow tools.
}

The rest of the paper is structured as follows. Section \ref{Research Methodology} describes the method used for this systematic literature review. Section \ref{results} presents the demographic information of the presented studies and answers the research questions of this SLR. The threats to validity are discussed in Section \ref{threats to validity}. \textcolor{black}{Section \ref{Related Work} provides a summary of the existing research.}
The insights into future research directions are discussed in Section \ref{discussion}. Finally, the conclusions of this study are presented in Section \ref{conclusion}.

\section{Research Methodology} \label{Research Methodology}
This Systematic Literature Review (SLR) aims to assess the existing research studies on recommending a software task to a software practitioner in a crowdsourcing platform. To conduct this SLR in an unbiased way, we adhered to the well-defined method outlined in Kitchenham and Charters' guidelines \cite{kitchenham2007guidelines}. The three primary stages of this research methodology are: 1) defining the review protocol, 2) conducting the review, and 3) reporting the review. The review protocol of this study consists of (i) research questions, (ii) search strategy, (iii) inclusion and exclusion criteria, (iv) study selection, and (v) data extraction and synthesis. The following subsections discuss these steps.

\subsection{Research questions} \label{research questions}

The research questions (RQs) were developed by the PICOC approach introduced by Petticrew and Roberts \cite{petticrew2008systematic}, explained in Kitchenham and Charters' guidelines \cite{kitchenham2007guidelines}. \textcolor{black}{This approach was originally developed in the field of Health Sciences but has been adapted for other domains, such as Information Technology. It helps streamline the identification of key elements essential for formulating clear research questions and selecting keywords for search string construction.} The PICOC (Population, Intervention, Comparison, Outcomes, Context) table for this SLR is available in Table \ref{table:1}. The intervention is the software technologies that address the issue, the population is the people affected by the intervention, the comparison is the technology used in software engineering that is being compared to the intervention, the outcomes are the factors of importance to practitioners, context refers to the context in which the comparison takes place, the people participating in the research, and the tasks being carried out.
\newline

\begin{table}[]
\caption{PICOC table for research questions}
\label{table:1}
\small
\begin{tabularx}{\columnwidth}{@{}ll@{}}
\toprule
Population  & Software practitioners in crowdsourcing platforms (e.g.; developers, testers) \\ \midrule
Intervention & Recommending software tasks in crowdsourcing platforms  \\ \midrule
Comparison  & N/A  \\ \midrule
Outcomes  & \begin{tabular}[c]{@{}l@{}}Existing software task recommendation methods in crowdsourcing /\\ Human factors that have been used when recommending tasks /\\ Crowdsourcing platforms and features that have been used to \\ recommend software tasks\end{tabular}  \\ \midrule
Context   & Crowdsourcing/Task recommendation/ Software Engineering practitioners  \\ \bottomrule
\end{tabularx}
\end{table}

\noindent\textbf{RQ 1: How has the task recommendation in crowdsourced software engineering been done so far and what is its current status?}
\newline 
\textcolor{black}{This research question is studied under four sub-questions mentioned below. It aims to investigate the state-of-the-art methods in task recommendation for CSE. It involves exploring the data sources and types of data extracted for task recommendations, categorizing the existing methods, and examining their advantages and limitations to understand the current landscape. It establishes the current state and methodologies of task recommendation systems, setting the context for subsequent RQs. Further, it lays the foundation for identifying opportunities for improvement.
\newline\textbf{RQ1.1: What data sources are used and what type of data are extracted to recommend tasks?}
\newline\textbf{RQ1.2: What are the existing software task recommendation methods?}
\newline\textbf{RQ1.3: What are the advantages of the existing task recommendation systems?}
\newline\textbf{RQ1.4: What are the limitations of the existing task recommendation systems?}}
\newline

\noindent\textbf{RQ2: What are the common types of tasks that are recommended to software practitioners? }
\newline
This RQ focuses on understanding the types of tasks commonly recommended to practitioners in CSE platforms. \textcolor{black}{It highlights the areas where task recommendation systems are currently concentrated and provides insights into the scope and diversity of tasks addressed by existing systems, offering a clearer picture of the industry’s focus. This RQ builds on RQ1 by narrowing the focus to the types of tasks addressed by these systems, highlighting industry priorities.}
\newline
\newline
\textbf{RQ 3: To what extent are human factors used to recommend CSE tasks?}
\newline
According to Dutra et al., human factors, also known as soft factors, human aspects, social factors, or non-technical factors, refer to the physical or cognitive features, or social behaviors of individuals \cite{dutra2021human}. Human factors cover a wide range of aspects, including personality, motivation, cognition, communication, decision-making, and cohesion among developers as they carry out their tasks \cite{lenberg2015behavioral}. Avison et al. \cite{avison1999action} state that ``Failure
to include human factors may explain some of the dissatisfaction with conventional information systems development methodologies; they do not address real organizations". Since software development is a human-centered process, human factors significantly impact both the process and its performance \cite{pirzadeh2010human}. \textcolor{black}{Given the human-centric nature of software development, understanding how human factors are integrated into CSE task recommendation systems helps identify gaps in addressing the social and cognitive needs of developers.} Hence, this RQ examines the human factors that are currently being considered in the software crowdsourcing task recommendation systems. It also identifies the human factors suggested to be included as future improvements in the selected primary studies. This RQ explores the integration of human factors, a critical dimension that complements the technical aspects addressed in RQ1 and RQ2.
\newline
\newline
\textbf{RQ4:What platforms are currently being used to recommend crowdsourced software
engineering tasks?}
\newline
This RQ identifies the different software crowdsourcing platforms currently being used for task recommendations. It further analyses the scale of the projects used to recommend tasks. \textcolor{black}{The answers to this research question provide insights into popular crowdsourcing platforms and their popularity and suitability for different task recommendation needs.}
\newline
\newline
\textbf{RQ5:Which features of platforms can be used to recommend crowdsourced software
engineering tasks?}
\newline
\textcolor{black}{This RQ identifies the features of crowdsourcing platforms that can be used to recommend tasks and the data items currently being extracted from them. The answers to this research question provide a comprehensive set of features and data attributes, offering actionable insights for system design and improvement in future studies.}

\subsection{Search strategy}

The search strategy used in this study was designed to retrieve as many relevant studies as possible \cite{kitchenham2007guidelines}, which is explained in detail in the following sub-sections.

\subsubsection{Search method}

The automatic and manual searches were carried out to find the relevant primary studies for this SLR. Scopus, IEEE Xplore, ACM Digital Library, Wiley, Science Direct, and Springer Link scientific databases were queried during the automatic search using the search query defined in Section \ref{Search terms}. Then, \textcolor{black}{backward and forward snowballing was done on the finalized relevant primary studies retrieved by the automatic search.}

\subsubsection{Search terms}  \label{Search terms}

The search query was formulated according to the Kitchenham and Charters’ guidelines \cite{kitchenham2007guidelines}. The PICOC approach (see Table \ref{table:1}) was used to select the key search terms. The search query was finalized after a series of pilot searches and examining the resulting studies. In some studies, it was observed that generic terms like “recommendation algorithm” and “recommendation system”  were used instead of any crowdsourcing-related terms within the title, abstract, and keywords, e.g., “good first issue recommendation system”. Hence, the aforementioned two terms were also added with the alternative search terms for “software crowdsourcing”. The final search query was formulated by connecting the alternative search terms with “OR” and “AND”, which are mentioned below. The alternative search terms used in the query are mentioned in Table \ref{table:2}. The wild cards were used except in the Science Direct database to cover a large number of alternative search terms. In Science Direct's advanced search, only 8 operators (e.g., AND/OR) are allowed per query. Hence, the combinations of the search terms were used to maintain consistency among the search queries used in different databases. Further, the search was not restricted by any particular period. \textcolor{black}{Finally, we randomly selected 8-10 papers from each digital database to confirm the credibility of our search string and validate that the included studies were the most pertinent to our review. }
\newline

\begin{table}[]
\caption{Alternative search terms used}
\label{table:2}
\small
\begin{tabular}{lp{0.75\textwidth}}
\toprule
\textbf{Main Search Term} & \textbf{Alternative Search Terms} \\
\midrule
Task Recommendation & task selection / issue recommendation /good first issue / feature recommendation / bug recommendation / bug report assignment / onboarding developer / repository recommendation / project recommendation / question recommendation / crowdtesting \\
\midrule
Software Crowdsourcing & software crowdsourcing / crowdsourc* software / github / open source / repository / stackoverflow / stack overflow / topcoder / software engineering / software develop* / open source development / recommendation algorithm / recommendation system \\
\bottomrule
\end{tabular}
\end{table}

\begin{figure}[h]
    \noindent\fbox{\begin{minipage}{1\textwidth}
        TITLE-ABS-KEY ( ( ``task recommendation" OR ``task selection" OR ``issue recommendation" OR ``good first issue" OR ``feature recommendation" OR ``bug recommendation" OR ``bug report assignment" OR ``onboarding developer" OR ``repository recommendation" OR ``project recommendation"  OR ``question recommendation" OR ``crowdtesting") AND ( ``software crowdsourcing" OR ``crowdsourc* software" OR ``github" OR ``open source" OR ``repository" OR ``stackoverflow" OR ``stack overflow" OR ``topcoder" OR ``software engineering" OR ``software develop*" OR ``open source development" OR ``recommendation algorithm" OR ``recommendation system" ) )
    \end{minipage}}
    \caption{Search query}
\end{figure}

\subsubsection{Data sources}

The search engines of scientific databases Scopus, IEEE Xplore, ACM Digital Library, Wiley, Science Direct, and Springer Link (Table \ref{table:3}) were used with the filters for automatic search. These are the main sources for potentially relevant research on software and software engineering \cite{chen2010towards}. Since the Springer Link database allows scanning only the title or the entire document, it returns a significant number of results for the query. Hence, only the top 300 results were analyzed. Since Google Scholar may return imprecise search results that may produce a large number of irrelevant results, and significantly overlap with ACM and IEEE on software engineering literature, it was not included in the data sources \cite{chen2010towards}.

\begin{table}[]
\caption{Queried scientific databases.}
\label{table:3}
\small
\begin{tabular}{p{0.2\textwidth}p{0.3\textwidth}p{0.4\textwidth}}
\toprule
\textbf{Database} & \textbf{Search Method} & \textbf{Filters Used} \\
\midrule
Scopus & Advanced search & TITLE-ABS-KEY \\
\midrule
IEEE Xplore & Command search & All Metadata \\
\midrule
ACM Digital Library & Advanced Search & Title, Abstract, Author Keyword filters added separately \\
\midrule
Wiley & Advanced search & Title, Keywords, Abstract filters added separately \\
\midrule
Science Direct & Advanced Search & Title, abstract or author-specified keywords. Does not support wildcards, hence removed the wildcards. Has 8 operators (AND/OR) per query limitation. Hence used combinations of the search terms. \\
\midrule
Springer Link & Advanced Search & Disciplines: Computer Science \\
\bottomrule
\end{tabular}
\end{table}

\subsection{Inclusion and exclusion criteria}

The inclusion and exclusion criteria applied to every study retrieved from digital libraries are shown in Table \ref{table:4} and Table \ref{table:5}, respectively. These criteria were refined during the search and paper filtering procedures to obtain an unbiased selection of studies. The finalized set of criteria was applied to each full-text article to choose the most relevant studies that align with the research questions and the aim of this SLR. All review papers and grey literature were excluded according to the SLR primary studies screening conventional procedure.

\begin{table}[]
\small
\caption{Inclusion criteria.}
\label{table:4}
\begin{tabularx}{\textwidth}{lX}
\toprule
\textbf{ID} & \textbf{Criteria} \\
\midrule
I01 & Peer-reviewed published studies that are available in full text \\
\midrule
I02 & Studies that are focused on recommending software engineering-related tasks to a software practitioner in a crowdsourcing platform \\
\midrule
I03 & \textcolor{black}{Study is published in one of the digital libraries we queried for this study}\\
\bottomrule
\end{tabularx}
\end{table}

\begin{table}[h]
\small
\caption{Exclusion criteria.}
\label{table:5}
\begin{tabularx}{\textwidth}{lX}
\toprule
\textbf{ID} & \textbf{Criteria} \\
\midrule
E01 & Unpublished and incomplete work, posters, and grey literature were excluded \\
\midrule
E02 & \textcolor{black}{ Studies that are focused on recommending a software practitioner to do a software engineering-related task published in a crowdsourcing platform}\\
\midrule
E03 & Papers that are not written in English language \\
\midrule
E04 & Short papers which have less than 4 pages \\
\midrule
E03 & \textcolor{black}{Papers that do not propose a recommendation system based solution }\\
\bottomrule
\end{tabularx}
\end{table}

\subsection{Study Selection} \label{Study Selection}

\begin{figure}
    \centering
    \includegraphics[width=1\linewidth]{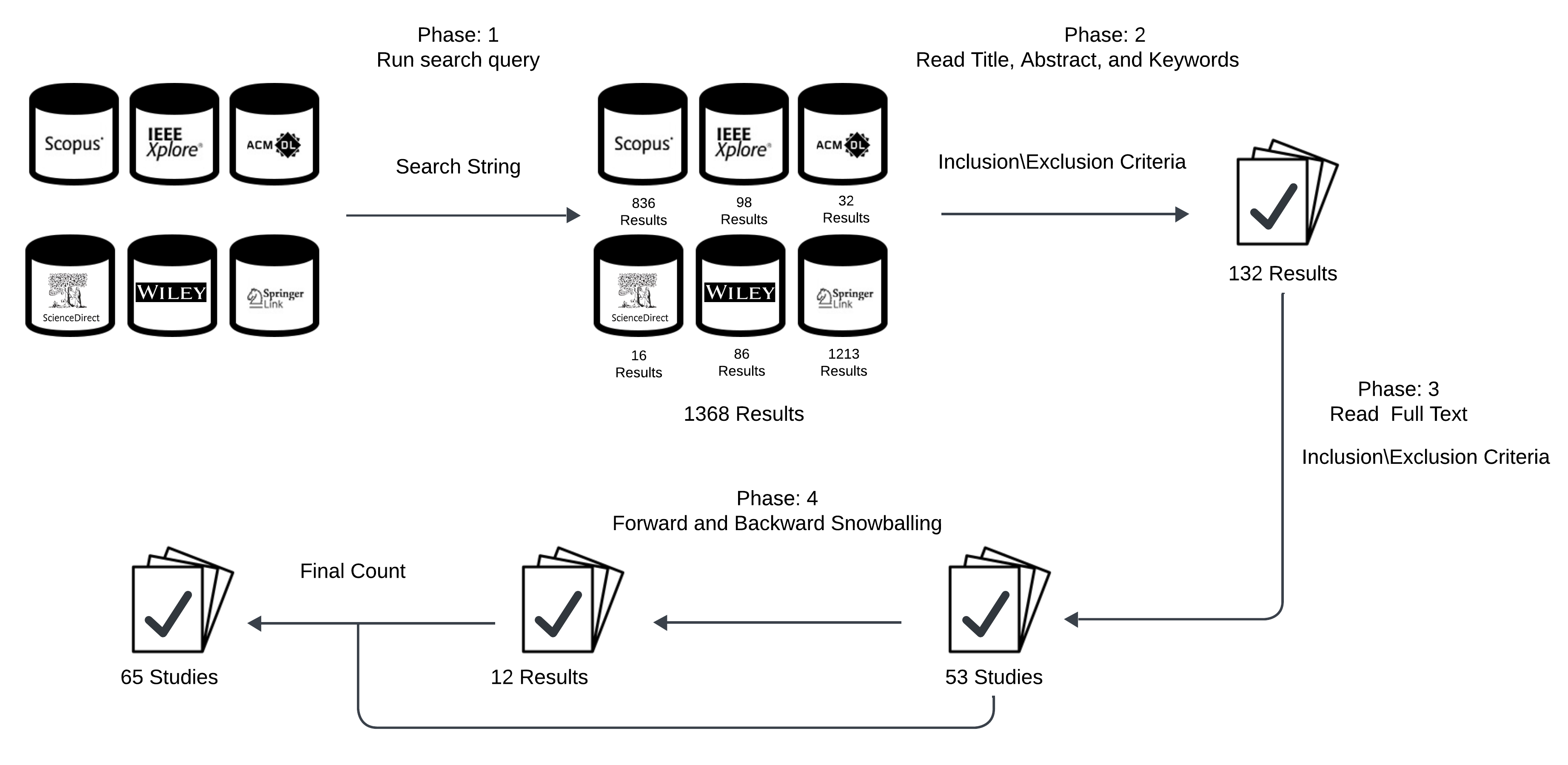}
    \caption{\textcolor{black}{Phases of study selection process}}
    \label{study selection diagram}
\end{figure}

The study selection was done in 4 phases. The breakdown of results for each phase according to the search method and library is available in Table \ref{table:6}. Figure \ref{study selection diagram} shows the number of studies retrieved at each phase of the study selection process.

\textbf{Phase 1}: The potentially relevant studies were retrieved from the chosen digital libraries using the designed search query. The total number of potential results for the query was 1368 (i.e., only the top 300 results were considered for the Springer Link database). Then, the inclusion and exclusion criteria were applied to the relevant studies set that had been retrieved.

\textbf{Phase 2}: During the second phase, the potential studies set was screened by reading the title, abstract, and keywords.  The papers which we had doubts regarding the relevancy were transferred to the next round of screening. The final count of potential studies at the end of the first screening round was 142.

\textbf{Phase 3}: During the third phase, the filtering was done by reading the full-text papers of the potentially relevant studies found during the second phase. \textcolor{black}{If the proposed solution of the study is not to recommend a software engineering-related task to a software practitioner or instead it is to recommend a software practitioner to do a task in a CSE platform, such studies were excluded in this stage after a careful evaluation of the proposed solution.} The data extraction of the primary studies was also done during this stage. After this filtering round, 51 papers were left as the primary studies.

\textbf{Phase 4}: The snowballing was done during the last phase for the finalized primary studies. Both backward and forward snowballing processes were used, and 9 and 12 potential studies were found, respectively. Then, the first filtration for those studies was done based on the title, abstract, and keywords. Then, 6 papers were left for backward snowballing, and 8 were left for forward snowballing at the end of filtration round 1. Then during the next filtration round, we read the full-text papers of those studies. After the second filtration, the total number of relevant primary studies finalized by backward snowballing was 5, and forward snowballing was 7. \textcolor{black}{The finalized set of primary studies are available in Appendix \ref{appendix:a}}.

\begin{table}[]
\small
\caption{Breakdown of results for the paper selection phases}
\label{table:6}
\begin{tabularx}{\textwidth}{XXXXX}
\toprule
& \textbf{Database} & \textbf{Initial Results Count} & \textcolor{black}{\textbf{Filter by Title, Abstract, and Keywords}} & \textcolor{black}{\textbf{Filter by Full text reading}} \\
\midrule
\multirow{6}{*}{Automatic Search} & Scopus & 836 & 85 & 6 \\
& IEEE Xplore & 98 & 26 & 20 \\
& ACM Digital Library & 32 & 19 & 15 \\
& Science Direct & 16 & 5 & 5 \\
& Wiley & 86 & 3 & 3 \\
& Springer & 1213 (top 300 analyzed) & 4 & 4 \\
\midrule
\multicolumn{2}{l}{\textbf{Total - Automatic Search}} & \textbf{1368} & \textbf{142} & \textbf{53} \\
\midrule
\multirow{2}{*}{Manual Search} & Backward Snowballing & 9 & 6 & 5 \\
& Forward Snowballing & 12 & 8 & 7 \\
\midrule
\multicolumn{2}{l}{\textbf{Total - Manual Search}} & \textbf{21} & \textbf{14} & \textbf{12} \\
\midrule
\multicolumn{4}{l}{\textbf{Final Paper Count}} & \textbf{65} \\
\bottomrule
\end{tabularx}
\end{table}

\subsection{\textcolor{black}{Quality assessment}}

\textcolor{black}{
We implemented a five-point scoring system to evaluate the quality of the selected primary studies based on five predefined questions. The scores range from very poor (1) and inadequate (2) to moderate (3), good (4), and excellent (5). This approach to quality assessment is widely used in SLRs \cite{naveed2024model,gunatilake2024impact}. Our quality assessment (QA) criteria were derived from Kitchenham’s guidelines \cite{kitchenham2007guidelines} and are outlined as follows:
\begin{itemize}
    \item QA1:Are the aims clearly stated?
    \item QA2: Are the measures used clearly defined?
    \item QA3: Is the solution clearly defined?
    \item QA4: Is there a clear outcome and results analysis reported?
    \item QA5: Are study limitations and possible future work adequately described?
\end{itemize}
}
\textcolor{black}{
The results of this assessment are presented in Appendix \ref{appendix:b}. Our assessment revealed that among the 65 included studies, 14 were of good quality, 39 were of average quality, and 12 were of poor quality. No studies were excluded based on the quality assessment score as recommending CSE tasks to practitioners is an emerging research area, and our selection was already constrained. We did not consider the citation count of each paper as a quality criterion to minimize publication bias and to ensure fairness toward recently published studies.
}

\subsection{Data extraction and synthesis}

\begin{figure}
    \centering
    \includegraphics[width=0.99\linewidth]{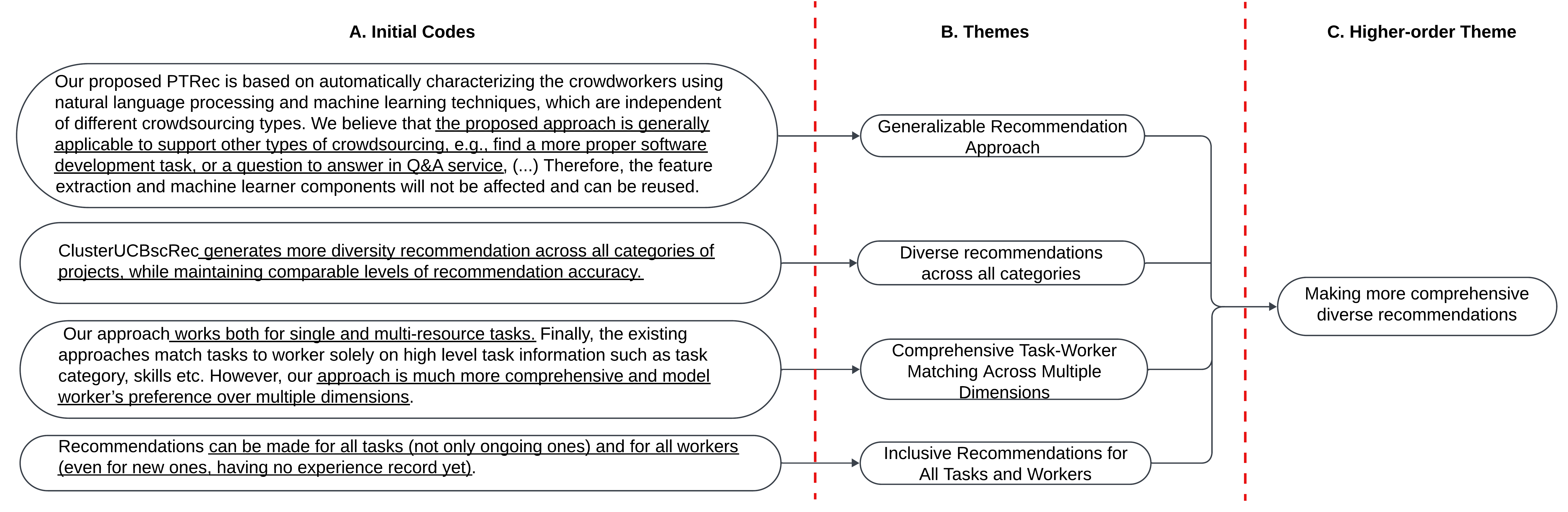}
    \caption{The steps of performing thematic analysis}
    \label{thematic_analysis}
\end{figure}

The data items extracted from the selected primary studies to answer the research questions of the SLR are presented in Table \ref{table:8}. The data items were retrieved and categorized based on each research question. The extracted data were collected in a Microsoft Excel Spreadsheet file. Data synthesis of the extracted data was done by thematic analysis \cite{shelby2008understanding,braun2006using}. This approach is often employed in systematic literature reviews (SLRs). It can be defined as an analysis of an extensive amount of structured, extracted data from individual studies to integrate and synthesise their main findings. The analysis was conducted by the first researcher, while the second researcher reviewed all extracted themes to validate them and identify any additional potential themes.
\textcolor{black}{We performed the thematic analysis using the five steps outlined by \cite{braun2006using}.
\begin{enumerate}
    \item Familiarizing with data: The extracted data items, such as task types (D08), advantages (D06), limitations (D07), data extracted from the user (D02), and data extracted from the task (D03), were carefully read and examined to develop an initial understanding.
    \item Generating initial codes: Lists of advantages, limitations, data extraction sources, etc., were compiled for each solution. In some cases, the original papers were revisited to ensure accuracy (Fig. \ref{thematic_analysis}.A).
    \item Searching for themes: The related initial codes from the previous step were grouped to form potential themes for each data item (Fig. \ref{thematic_analysis}.B).
    \item Reviewing themes: Identified themes were compared and analyzed, assessing whether some needed to be merged or removed.
    \item Defining and naming themes: Finally, clear and meaningful names were assigned to each identified data extraction source, advantage, limitation, etc., ensuring they accurately represented the findings (Fig. \ref{thematic_analysis}.C).  
\end{enumerate}
}

\textcolor{black}{Figure \ref{thematic_analysis} shows the application of thematic analysis to analyze various advantages and identify the "Making more comprehensive diverse recommendations" theme. }

\begin{table}[]
\caption{Data items extracted}
\label{table:8}
\resizebox{\textwidth}{!}{%
\begin{tabular}{@{}llll@{}}
\toprule
    & Data Item                        & Description                                                                   & RQ               \\ \midrule
D01 & Title                            & N/A                                                                           & RQ1              \\
D02 & Type of data extracted from user & What kind of data extracted from the user (eg: preferences)                   & RQ1              \\
D03 & Type of data extracted from task & What kind of data extracted from the task (eg: skills required) & RQ1 \\
D04 & Method steps                     & Implementation steps of the proposed system                                   & RQ1              \\
D05 & Method type                      & The algorithmic category of the proposed system (eg: Matrix factorization)    & RQ1              \\
D06 & Advantages                       & Advantages of the proposed system                                             & RQ1              \\
D07 & Limitations and future work      & Limitations and the future work of the study                                  & RQ1              \\
D08 & Task type                        & Type of the task recommended                                                  & RQ2              \\
D09 & Human factors used               & Human factors included in the study                                           & RQ3              \\
D10 & Human factors suggested          & Human factors suggested to be included as a future work of the study          & RQ3              \\
D11 & Features                         & Features of the crowdsourcing platform used to extract data (eg: ReadMe file) & RQ4              \\
D12 & Extracted data                   & Data extracted from the features (eg: user activities like star, fork)        & RQ4              \\
D13 & Platform                         & Name of the crowdsourcing platform used in the study (eg: Github)             & RQ5              \\
D14 & Project name(s)                  & Name(s) of the projects that were used in the study                           & RQ5              \\
D15 & Comments                         & Additional information                                                        & N/A              \\
D16 & Author(s)                        & Author(s) names                                                               & Demographic data \\
D17 & Year                             & Publication year                                                              & Demographic data \\
D18 & Venue                            & Publication venue                                                             & Demographic data \\ \bottomrule
\end{tabular}%
}
\end{table}

\section{Results}\label{results}

The analysis and synthesis of the data extracted from our finalized \textcolor{black}{65} primary studies are reported in the following subsections. The results discussed in this section are derived by synthesizing the data with minimal interpretations directly extracted from the selected primary study set.

\subsection{Demographic Data}
\begin{figure}
    \centering
    \includegraphics[width=1\linewidth]{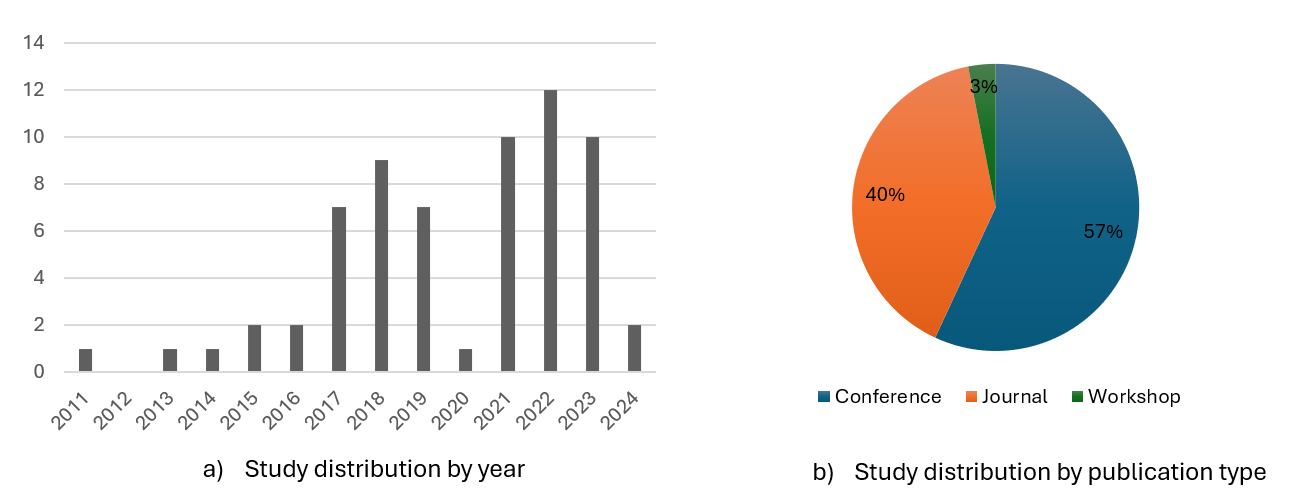}
    \caption{\textcolor{black}{Publication trends}}
    \label{Publication trends}
\end{figure}

Of the \textcolor{black}{65} primary studies selected for this SLR, \textcolor{black}{37 (56.9\%)} were conference papers, \textcolor{black}{26 (40\%)} were journal papers, and 2 (3.1\%) were workshop papers. Figure \ref{Publication trends} shows the yearly distribution and publication type distribution of the selected papers. The selected primary studies were published between 2011 and \textcolor{black}{2024} era even though we did not limit our search to a specific period. None of the included studies were published in 2012. There is an overall increase in the number of publications from 2011 to 2018. Then, in 2019 and 2020, there was a drop in the number of publications. The COVID-19 outbreak might have caused the sudden publication drop seen in 2020. However, it is not certain. Overall, \textcolor{black}{52\% }of the selected primary studies set were published from 2021 to \textcolor{black}{2024}.

\subsection{RQ1: How has the task recommendation in crowdsourced software engineering been done so far and what is its current status?}

This RQ aims to analyze how software task recommendations in crowdsourcing platforms have been made to date. To answer this RQ, we analyzed the data extracted to recommend tasks, proposed methods, the advantages, and the limitations of the proposed methods. The results of the analysis are discussed in the following sub-sections.

\subsubsection{RQ1.1: What data sources are used and what type of data are extracted to recommend tasks?} \label{extracted data}

After analyzing the primary studies, we identified that the data used to recommend tasks in the selected primary studies were sourced from 4 sources, described below. Table \ref{table:7} shows the extracted data according to the data acquisition source.
\begin{enumerate}
\item \textbf{Software Practitioner's Profile}: One of the main sources of data is the software practitioner's profile. Such profile is mainly used to extract the data listed below:
\begin{itemize}
    \item \emph{Skills} -  Practitioner's skills, such as the programming languages worked with, tools, and platforms familiar with, were extracted from their profiles. For example, Study S14 extracted software practitioner's skills in programming languages, frameworks, platforms, etc from the practitioner's profile.
    \item  \emph{Ratings} - Ratings received for previously completed tasks reflect the expertise level of the practitioner. For example, Study S29 extracted ratings received for previously completed tasks as ratings capture how well a worker has worked on other criteria in the past.
    \item \emph{Workload} - It is important to evenly distribute the newly reported bugs among potential developers who are not overloaded by work. Study S48 extracted the developer's work history from the developer's profile to decide on the current workload.
\end{itemize}

\item \textbf{Task or Project}: The task or project itself provides insights about the job required to be completed. Data listed below are extracted from the task or project itself.
\begin{itemize}
    \item \emph{Skills, expertise} - The skills and expertise required to complete the task are popular data types available for the tasks or projects. Study S5 and S27 extracted the programming language from the task to determine the required skills to complete the task. 
    \item \emph{Task’s competitive advantage, Technologies hotness} - Study S2 captures the task’s competitive advantage over the other open tasks. Technologies hotness refers to the relative level of demand and supply of a particular technology. Study S39 defines the hotness of technology on a five-point scale from 1 (most popular) to 5 (least popular) for the technologies mentioned in the task requirements.
    \item \emph{Progress, Status, Severity} - These data items provide quick insights into the task. The status of a task is often described using different tags as `fixed', `invalid', `duplicate', `wontfix', and `worksforme' (e.g., S42 extracted `fixed' or `verified' status from the task). The severity describes the importance using `critical', `major', etc tags. (e.g., S48 extracted the severity from the bug reports)
    \item \emph{Resources} - The resources allocated for the task like the time and budget are extracted from the task. For example, Study S15 captured the budget and estimated completion time. Study S28 extracted the task hours and rates from the task poster.
    \item \emph{Domain} - The domain refers to the specific area the task or project focuses on. Study S15 extracted project category and type from the posted task.
    \item \emph{Code involved} - The code involved in the required fix is another data extracted from the task. Study S7 used the source code when building the feature vector of keywords used to recommend the task. 
    \item \emph{Popularity} - The task or project popularity is another popular data extracted. Study S17 extracted the number of stars received and the number of forks which provides insights about the popularity of the repository among the community.
    \item \emph{Feedback} - online feedback of the project sponsors and developers on the task was extracted in Study S15.
    \item  \emph{Task description sentiments} - Study S9 extracted the sentiments associated with task description such as negative and positive emotions as they impact newcomers selecting which issues to work on.
    \item \emph{Contributor’s social characteristics} - Social characteristics of the contributors, such as influence gained from developer's friends to bid for similar tasks, and the frequency of interaction with other users were gathered from the task or the project itself. For example, In Study S8, inactive developers received recommendations by pooling together task recommendations from their friends who bid on tasks.
\end{itemize}

\item \textbf{Previous Contributions}: Some studies extracted data from the software practitioner's previously contributed tasks or projects as they provide insights into the practitioner's preferences, skills, expertise, etc. 
\begin{itemize}
    \item \emph{Skills, Expertise} - Skills and expertise level required to complete the task were extracted from previous contributions to gather data about the practitioner's past skills and expertise. For example, in Study S2, expertise is considered as a crowd tester's capability of detecting bugs. Hence, it is extracted based on the number of bug reports submitted by the worker for each domain.
    \item \emph{Activeness} - Study S2 extracted the number of bugs or reports submitted within a specified period to measure the activeness of testers.
    \item \emph{Historical behavior sequences} -  Historical behavior sequences refer to creating a pull request, starring, forking, watching a repository, commenting, bidding for a task, etc. These are considered reflections of the developer's programming preferences or interest in the project. For example, Study S17 extracted `Create', `Fork', `Star', and `Pull' Request activity data from the developer's previously contributed projects.
    \item \emph{Sentiments exhibited in communication } - Study S59  proposes that newcomers exhibiting a more positive communication style are more likely to receive assistance from experienced developers when dealing with complex issues. Further, Newcomers with less coding experience and who express negative sentiments than their peers tend to prefer adaptive issues over corrective or perfective ones. Those who contributed to the first issues in web library and framework projects generally display more positive sentiment characteristics. Hence, they extracted mean and median textual sentiment polarity values of previously contributed Pull requests and reported issues.
    \item \emph{Task intensity or complexity, Severity } - Task intensity or complexity (strong, weak, etc.), severity (critical, major, normal, etc.) were also extracted from previous contributions. Study S10, extracted intensity (strong, medium, weak, etc.) and severity data (critical, major, etc.).
    \item \emph{Contributed source files or components} - Source files refer to the individual code files that are related to a specific feature, while components are distinct modules that group related issues and functionalities within the project. Study S3, extracted the previously contributed components of the software projects by the developer.
    \item \emph{User role evolution} - According to Study S49, each user can take on the roles of both asker and answerer at the same time in question answering crowdsourcing platforms. However, user roles vary over time. Hence, they extracted the software practitioner's simultaneous role evolution as a question-asker and answerer in the crowdsourcing platform.
\end{itemize}

\item \textbf{Direct Data Collection}: In some cases, a customized platform is introduced as a website to directly gather the required specific data, listed below, from the software practitioners. The main reasons for introducing these platforms to gather data include: existing crowdsourcing platforms not storing some required data (e.g., personality types) and none of the platforms being open to providing information on their developers or clients \cite{tunio2017impact}\cite{gilal2022task}.

\begin{itemize}
    \item \emph{Demographic data} - Demographic data such as education level, location, etc were gathered by a new website in Study S4. 
    \item \emph{Personality type} - Studies S4 and S57  gathered personality types of the developers.
    \item \emph{project and skill preferences}- Study S1 introduced a website that enables users to select a project and input their skill set to receive personalized recommendations.
\end{itemize}
\end{enumerate}

\begin{center}
\begin{tcolorbox}[colback=gray!5!white,colframe=black!75!black]
\textbf{RQ1.1 Summary:} The data used to recommend tasks in the selected primary studies were sourced from 4 sources: software practitioner's profile, task or project itself, previously contributed tasks or projects available in the crowdsourcing platform, and direct data collection via a customized platform. The most popular data type extracted from these sources is the software practitioner’s historical behavior sequences, such as starring, forking, and watching a repository.
\end{tcolorbox}
\end{center}

\begin{longtable}{@{}p{0.3\textwidth}p{0.4\textwidth}p{0.2\textwidth}ll@{}}
\caption{Extracted data according to the data acquisition source}
\label{table:7} \\
\toprule
\textbf{Data Acquisition Source} & \textbf{Gathered Data} & \textbf{Studies} \\
\midrule
\endfirsthead
\multicolumn{5}{c}%
{{\tablename\ \thetable{} -- continued from previous page}} \\
\toprule
\textbf{Data Acquisition Source} & \textbf{Gathered Data} & \textbf{Studies} \\
\midrule
\endhead
\midrule
\multicolumn{5}{r}{{Continued on next page}} \\
\endfoot
\bottomrule
\endlastfoot
\multirow{4}{*}{Software Practitioner's Profile} & Skills & S14, S15, S16, S26, S27, S29, S46, S48, S53, S59 \\
& Ratings & S29 \\
& Workload & S48 \\
\midrule
\multirow{10}{*}{Task or Project} & Skills, Expertise & S1, S5, S12, S15, S27, S34, S52, S55, S62 \\
& Task's competitive advantage, Technologies hotness & S2, S39 \\
& Progress, Status, Severity & S2, S3, S10, S42, S43, S48 \\
& Resources (time, money) & S4, S15, S28 \\
& Domain & S4, S10, S14, S15, S23, S26, S30, S31, S35, S36, S40, S46, S54, S57, S58, S59 \\
& Code involved & S7, S41, S50 \\
& Popularity & S14, S17, S18, S26, S36, S51, S56 \\
& Feedback & S15 \\
& Task description sentiments & S9 \\
& Contributor's social characteristics & S8, S12, S34, S36, S38, S53, \textcolor{black}{S65} \\
\midrule
\multirow{9}{*}{Previous Contributions} & Skills, Expertise & S2, S5, S10, S25, S27, S28, S29, S32, S37, S39, S40, S41, S42, S43, S45, S46, S52, S55, \textcolor{black}{ S64, S65} \\
& Activeness & S2 \\
& Historical behavior sequences & S5, S7, S8, S10, S17, S18, S19, S20, S21, S22, S23, S24, S32, S36, S47, S50, S52, S55, S60, \textcolor{black}{S65} \\
& Sentiments exhibited in communication & S59 \\
& Task intensity or complexity, Severity & S10, S33 \\
& Contributed source files or components & S3, S13, S48 \\
& User role evolution & S49 \\
\midrule
\multirow{3}{*}{Direct Data Collection} & Demographic data & S4 \\
& Personality types & S4, S57 \\
& Project and skill preferences & S1 \\
\end{longtable}

\subsubsection{RQ 1.2: What are the existing software task recommendation methods?} \label{recommendation methodologies}

To categorize the existing task recommendation methods of the primary studies, we used the taxonomy introduced in \cite{roy2022systematic}. Hence, the recommendation systems in the primary studies were broadly categorized into \emph{content-based}, \emph{collaborative filtering}, and \emph{hybrid recommendation systems} categories. 

\textbf{Content-based}. In \emph{content-based} recommendation systems, all data items are gathered and categorized into distinct item profiles according to their descriptions or features \cite{roy2022systematic}.  For example, Study S3 created a user profile for each developer based on the previously fixed issues. Whenever a new issue arrived, they compared the  30 most frequent words or tokens of the developer's profile with TF-IDF (Term Frequency - Inverse Document Frequency) values of all tokens of the new issue’s preprocessed title and description to determine if the new issue falls in the developer's preferred task category. Content-based approaches are the most popular category of the selected primary study set (65\% of the primary study approaches). Further, to sub-categorize the content-based systems, we used the taxonomy introduced in \cite{taghavi2018new}. Hence, the content-based category was further subcategorized into the following sub-categories:
\begin{itemize}
    \item \emph{Machine Learning}: Deciding whether to recommend an item to a user can be viewed as the problem of modeling a task and predicting if the user will like it. To achieve this, different standard machine learning techniques can be used to predict if a user will be interested in an item or not \cite{taghavi2018new}. For example, Study S2  extracted 60 features automatically and employed a random forest learner to generate dynamic and personalized task recommendations aligned with a worker's skills and preferences. These features encompass the task's progress status, crowdworkers' availability, the matching degree between the task and crowdworkers, as well as the task’s competitive advantages among other open tasks. From the content-based approaches, 56.4\% are machine learning based approaches. We further categorized the selected primary studies into the below machine learning categories mentioned in the taxonomy introduced in \cite{taghavi2018new}.
    \begin{itemize}
        \item \emph{Classification} - Classification algorithms learn decision boundaries and separate
instances in a multi-dimensional space \cite{taghavi2018new}. For example, Study S1 used the Random Forest Classifier to determine relevant domain labels for the issues based on the inputted issue text.
        \item \emph{Artifical Neural Networks (ANN)} - Artificial Neural Networks (ANNs) are models inspired by the behavior of biological neurons. They attempt to mimic the brain's way of processing information and learning to some extent  \cite{taghavi2018new}. Study S17 used a multi-layer graph convolutional network to recommend repositories.
        \item \emph{Reinforcement Learning} - Reinforcement learning is the problem faced by an agent that learns behavior through trial-and-error interactions with a dynamic environment \cite{kaelbling1996reinforcement}. The machine learning category was further extended by us by adding the reinforcement learning category to the existing taxonomy. Study S15 introduced a clustering-based reinforcement learning solution to recommend tasks to newcomers.
    \end{itemize}
     
     \item \emph{Vector-based Representation}: This is based on the idea that a simple way to describe an item is by listing its specific attributes or features. When user preferences are expressed based on these features, the recommendation task involves matching them accordingly  \cite{taghavi2018new}. For example, Study S11 recommended similar bug reports by calculating three similarity scores based on the bug titles and descriptions, product, and component information for the query bug and each of the pending bugs. In the final step, they merged the three similarity scores to generate a final score for each pending bug and recommended the bugs with the highest final scores as the ones most similar to the query bug. From the content-based approaches, 43.6\% are vector-based representations.
\end{itemize}
 
\textbf{Collaborative Filtering}. These approaches rely on determining the similarity between users \cite{roy2022systematic}. This method begins by identifying a group of users, denoted as X, whose preferences, interests, and dislikes closely resemble those of user A. 20\% of primary studies are collaborative filtering approaches. Further, these methods can be categorized as follows:
\begin{itemize}
    \item \emph{Model-based}: Model-based systems use gathered information to build a model that generates recommendations  \cite{taghavi2018new}. They employ various data mining and machine learning algorithms to create a model that predicts a user's rating for an unrated item. Instead of relying on the entire dataset to generate recommendations, these systems extract features from the dataset to build the model \cite{roy2022systematic}. For example, Study S8 built a developer social network and then used singular value decomposition (SVD) to recommend tasks to active developers. Of the collaborative filtering-based primary study approaches, 75 \% are model-based.
    \item \emph{Memory-based}: Memory-based approaches recommend new items based on the preferences of the neighborhood. They mostly use similarity metrics to measure the distance between two users or two items \cite{taghavi2018new}. These methods directly utilize a utility matrix for making predictions. The first step involves constructing a model, which is essentially a function that uses the utility matrix as input. Recommendations are then generated through a function that takes both the model and the user's profile as input \cite{roy2022systematic}. For example, Study S31 computed a similarity matrix for test tasks using past test tasks completed by all crowdworkers. Then, based on this similarity matrix and a crowdworker’s history of test tasks, their method determines the crowdworker’s preference for other test tasks for recommendations. 25 \% of the collaborative filtering-based primary study approaches are memory-based.
\end{itemize}

\begin{figure}
    \centering
    \includegraphics[width=1\linewidth]{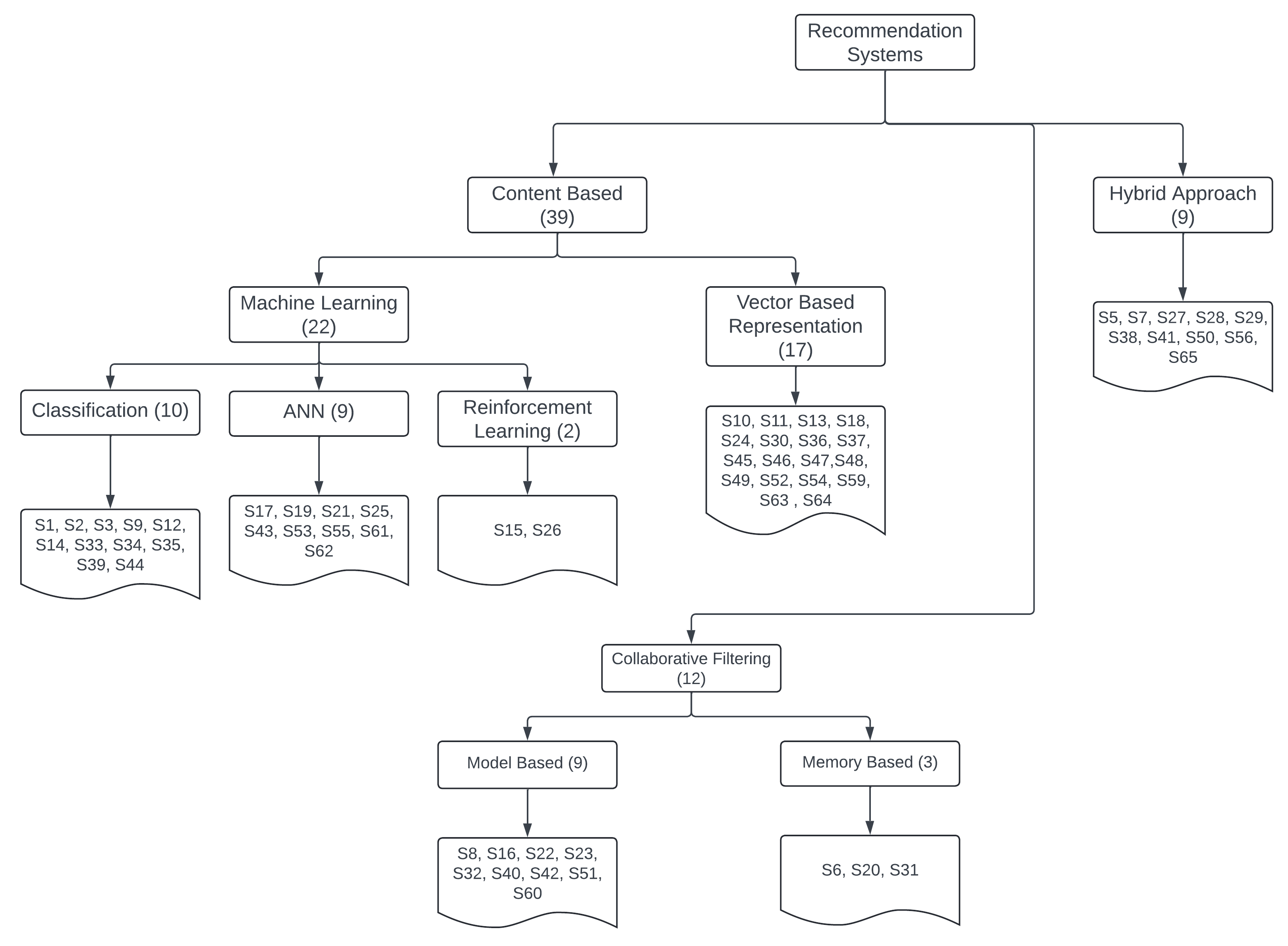}
    \caption{\textcolor{black}{Taxonomy for proposed methodologies}}
    \label{fig:2}
\end{figure}

\textbf{Hybrid Approach}. \emph{Hybrid methods} might integrate outcomes obtained from distinct methods, it could apply content-based filtering alongside collaborative methods, or use collaborative filtering within a content-based approach \cite{roy2022systematic}. As an example, Study S27 is a combination of developer profiles, topic distributions of the crowdsourcing tasks (content-based filtering), and a social network to measure a developer’s influences (collaborative filtering). 15\% of the primary studies use hybrid approaches. Figure \ref{fig:2} shows the extended taxonomy for the recommendation systems introduced in the primary studies.

\begin{center}
\begin{tcolorbox}[colback=gray!5!white,colframe=black!75!black]
\textbf{RQ1.2 Summary:} The existing software task recommendation methods can be categorized as content-based, collaborative filtering, and hybrid approaches. Content-based approaches are the most popular category, accounting for 65\% of the approaches. Collaborative filtering-based approaches come in second by 20\%, followed by hybrid approaches by 15\%.
\end{tcolorbox}
\end{center}

\subsubsection{RQ1.3: What are the advantages of the existing task recommendation systems?}

After analyzing the advantages of each primary study (i.e., data item D06), we identified three main categories of advantages. These categories are discussed in detail in the following subsections.
\newline
\newline
\textbf{Increasing performance, accuracy, and optimizing solutions}: The performance, accuracy, and system optimizations in the selected primary studies were mainly achieved by using two or more characteristics mentioned in Section \ref{extracted data}. For example, studies S8, S12, S34, S36, S38, S53, and \textcolor{black}{S65} used software practitioner's social connections with fellow software practitioners along with other characteristics to improve the accuracy of the systems. According to Study S12, the use of social metrics in their recommendation system improved the precision of the used models up to 15.82\%. Further, some studies (e.g., S5, S7, S27) used hybrid recommendation systems (mentioned in section \ref{recommendation methodologies}) to mitigate the limitations that occurred when using one recommendation approach.
\newline
\newline
\textbf{Making more comprehensive diverse recommendations}: Making more comprehensive diverse recommendations involves offering a wide range of recommendations that cater to different platforms, activities, or user preferences. For example, Study S2 is based on the automatic characterization of crowd workers through the use of machine learning and natural language processing techniques. Hence, those are not dependent on the specific types of crowdsourcing. Even though the system is focused on crowd testing, the proposed method can be used to support other crowdsourcing activities such as finding a better software development task or a question to answer in a Q\&A platform.
\par
The recommendation system proposed in S15 can make diverse recommendations across different project categories. The proposed system achieves it by analyzing feedback from users and behaviors which results in learning the preferred projects and developers' skill levels.
The recommendation system proposed in study S28 can make more comprehensive recommendations as it models workers’ preferences over multiple dimensions. In Study S39, the recommendations can be made to all tasks, not just those that are currently in progress, and to every employee, including those who are new and lack prior experience. The system analyses worker's preference patterns, technologies' hotness, and the projection of winning chances to make recommendations.
\newline
\newline
\textbf{Providing recommendations to newcomers ( “cold start” users)}: The “cold start” problem refers to the challenge of providing recommendations to new users who don't have enough historical data or activity to generate accurate recommendations. Some of the existing literature can provide recommendations to cold-start users. Study S8 addresses the  “cold start” problem by integrating the recommended tasks of their taskbidding friends. Study S9, automatically identifies issues that newcomers can fix by analyzing the history of issues that have been resolved by using the title and description of issues. The method proposed in S14 uses a topic sampling-based genetic algorithm to recommend tasks to newcomers and transforms the cold-start developer recommendation problem into a multi-optimization problem. Study S15 addresses the  “cold start” problem by learning about project preferences and development skill levels of the software practitioner through analyzing and mining user feedback and behaviors. 

Study S28 solves the  “cold start” problem by modelling worker’s preferences over multiple dimensions. Study S44 recommends tasks to newcomers by integrating repository history data  (such as the number of prior commits made by the issue reporter at the time of the issue's creation) and global GitHub profiles for all issue reporters and repository owners (e.g., stars received, commits). Study S59 recommends tasks to cold-start users by analyzing possible features and the characteristics of the newcomers’ chosen first issues. A new cross-platform recommendation approach for software crowdsourcing was introduced in S61 by transferring data and knowledge from other popular software crowdsourcing platforms. Study S61 adopts a transfer learning approach to solve the platform  “cold start” problem.

\begin{center}
\begin{tcolorbox}[colback=gray!5!white,colframe=black!75!black]
\textbf{RQ1.3 Summary:} There are several advantages in the existing CSE task recommendation systems, such as increased performance, accuracy, and optimized solutions, making more comprehensive and diverse recommendations, and providing recommendations to newcomers (“cold start” users).
\end{tcolorbox}
\end{center}

\subsubsection{RQ1.4: What are the limitations of the existing task recommendation systems?}
\label{existing imitations}
We analyzed the limitations pointed out in the primary studies and categorized those limitations into three main categories: dataset, approach, and evaluation. There were no any mentioned limitations in Studies S5, S6, S7, S14, S19, S22, S25, S29, S41, S46, S57, S58, S60 and S63. The limitations are discussed in detail in the following sub-sections.
\begin{enumerate}[label=(\roman*)]
\item \textbf{Limitations in the datasets (41 papers):}
In several primary studies, the dataset used in the primary studies has limitations. Due to the nature of the available information on crowdsourcing platforms, some factors that could be considered for recommendations were ignored. This limitation can lead to inadequate or biased recommendations, as crucial factors influencing recommendation quality, such as the difficulty of tasks or the activity levels of developers, may not be captured. For example, no information indicates task difficulty/ease level in the dataset used in Study S2. Study S53 could not extract several features that affect developers' project selection, like the mailing list and download count, due to the nature of the data they obtained from GitHub. Making recommendations for inactive developers is challenging due to the relative scarcity of available information about them. As a result, the proposed metrics in Study S20 need to be evaluated using other datasets. Consequently, the recommendations may lack accuracy and relevance, potentially resulting in the inability to recommend the most suitable task for the software practitioner. The data set used in Study S39 was extracted from Topcoder. Hence, the conclusions they made might not be generalizable to other platforms.
\item \textbf{Limitations in the approaches (7 papers):}
One of the key limitations identified in the proposed approaches is the lack of generalizability of the solution. The recommendation model in Study S3 primarily follows a content-based approach, limiting its ability to suggest issues to developers based solely on past contributions. It lacks the capability to provide more insightful recommendations that consider serendipity aspects across various components of the same product. Hence, the recommendations could be narrow. The NLP suggestions used in Study S12 might not work as expected in other languages like C++. Therefore, the approach cannot be utilized in projects using different programming languages. Whereas in Study S9, the projects within the scope of the study exhibit diverse practices, backgrounds, resources, and goals. These factors are not adequately considered during the model-building process. Hence, the recommendation system proposed in the study may not be applicable or valid across a broader range of projects with different contexts. Another key limitation is that some systems cannot recommend tasks to newcomers ( “cold start” problem). In Studies S2, S14, S15, S17 and \textcolor{black}{S65}, the proposed recommendation systems cannot recommend tasks to newcomers. As mentioned earlier in section \ref{research questions}, human factors play a major role in the software engineering context. However, only 24 primary studies have considered human factors in their recommendation systems. These human factors are analyzed in section \ref{human factors rq}. 39 primary studies ignored human factors in their recommendation systems. Some primary studies that  ignored human factors are, S3, S6, S7, S10, S11, S14, S17, S20, 21, S22, etc.

\item \textbf{Limitations in the evaluation process (4 papers):} Study S12 restricted predictions for issues with linked Pull Requests (PRs). Hence, it may have inadvertently introduced bias and overlooked significant issues that lacked associated PRs. This approach could have led to incomplete or skewed ground truth data, thereby affecting the accuracy and reliability of their prediction model. Study S13 only used the Spring Framework as the experimental project, and selecting only 15 developers as subjects might limit the generalizability of the proposed approach. This approach risked neglecting the diverse practices and contexts prevalent in other software projects, potentially undermining the relevance and applicability of the recommendations in real-world scenarios. Study S26's reliance on users with specific contribution thresholds and reputation scores (users who have made at least more than or equal to 5 contributions and a reputation score above 10) could have overlooked valuable insights from less active or newer contributors, leading to a biased understanding of users. Hence, the work is limited to the users they studied.

\end{enumerate}

\begin{center}
\begin{tcolorbox}[colback=gray!5!white,colframe=black!75!black]
\textbf{RQ1.4 Summary:} There are three main limitation areas in the existing studies: limitations in the dataset, limitations in the approach, and limitations in the evaluation. Limitations in the dataset include overlooking crucial factors like task difficulty, potentially leading to biased recommendations. Proposed approaches often lack generalizability, such as content-based models providing narrow suggestions or NLP methods being programming language-specific. Limitations in the evaluation may introduce bias or limit generalizability, like focusing on specific project contexts or user thresholds, undermining the relevance of findings. These limitations underscore the challenges in developing robust and comprehensive recommendation systems for software practitioners.
\end{tcolorbox}
\end{center}

\subsection{RQ2: What are the common types of tasks that are recommended to software practitioners?}

Table \ref{tasks to recommend} shows the types of tasks recommended to software practitioners in crowdsourcing platforms.
\begin{itemize}
    \item \emph{Repository to contribute}: It refers to recommending a software repository where practitioners can make different contributions such as fixing bugs, adding features, or making improvements. 42.8\% of the selected primary studies are to recommend a repository that the software practitioner can contribute to.
    \item \emph{Fixing issues}: Recommending an issue that practitioners can fix is another type of task recommended to the practitioners. 21.4\% of the tasks in primary studies are to recommend issues to fix.
    \item \emph{Testing tasks}: It refers to recommending a testing task to a software tester.
    6.6\% of the tasks in primary studies are to recommend software testing tasks.
    \item \emph{Good First Issues (GFIs)}: Another type of task is called GFIs, i.e., beginner-friendly issues within software projects that newcomers can tackle as their first contributions. 6.6\% of the tasks in primary studies are to recommend GFIs.
    \item \emph{Question to answer}: It refers to recommending a question from a Q\&A platform to answer. 6.6\% of the tasks in primary studies are to recommend questions.
\end{itemize}
 
\begin{center}
\begin{tcolorbox}[colback=gray!5!white,colframe=black!75!black]
\textbf{RQ2 Summary:} Recommending a repository that the software practitioner can contribute to, tasks to fix issues, testing tasks, Good First Issues (GFI), and questions to answer are the most common task types recommended in the existing literature. 42.8\%  of the selected primary studies focus on recommending a repository to contribute.
\end{tcolorbox}
\end{center}

\begin{table}[]
\caption{Types of tasks recommended in the existing systems}
\label{tasks to recommend}
\small
\centering
\begin{tabularx}{\textwidth}{lXl}
\toprule
\textbf{Task Type} & \textbf{Studies} & \textbf{Count} \\
\midrule
Repository to contribute & S5, S6, S7, S15, S16, S17, S18, S19, S20, S21, S22, S23, S24, S30, S32, S36, S47, S50, S51, S52, S53, S54, S55, S56, S61, S62, S63, \textcolor{black}{S64, S65} & 27 \\
\midrule
Fixing issues & S1, S10, S11, S12, S34, S37, S41, S42, S43, S44, S45, S48 & 12 \\
\midrule
Testing tasks & S2, S31, S58, S60 & 4 \\
\midrule
Good First Issues & S9, S33, S35, S59 & 4 \\
\midrule
Question to answer & S25, S26, S38, S49 & 4 \\
\bottomrule
\end{tabularx}
\end{table}

\subsection{RQ3: To what extent are human factors used to recommend CSE tasks?} \label{human factors rq}

\textcolor{black}{Human factors encompass personality traits, social and behavioral preferences, and sentiments, which are intrinsic to individuals and require different methods of measurement and analysis.}
To answer this research question, we analyzed the human factors (i.e., D09 and D10 in \ref{table:7}) currently used in the existing recommendation systems and the human factors suggested to be included as future improvements to the recommendation systems. Table \ref{human factors} shows these human factors.

\emph{Topic and task type preference} is used in 9 studies. It is the most used human factor in the existing studies. In study S2, task preference measures the extent to which a prospective task might capture the interest of a potential crowdworker. The worker's willingness and potential to detect bugs increase with their preference of taking up a task. 
Study S35 will incorporate the preferences of the newcomers as future improvements to their recommendation system.

\emph{Skill preference} is the second most used human factor used in 6 studies. For example, Study S1 allows a user to select a project to work on, enter their skill set (in terms of predefined categories), and get a list of open issues that would need those skills.

\emph{Social and behavioral preferences} are another preference type considered in the existing studies. For example, Study S8 introduced a social influence based method to recommend tasks. Study S5 suggests using the social connections of software practitioners as future improvements to their recommendation system.

Studies S4, S57, and S58 investigated the impact of \emph{personality} on task selection in crowdsourcing software development. In study S4, an empirical experiment was carried out to assess the impact of personality on task selection based on the three key components of a task: type, money, and time \cite{tunio2017impact}. Study S57 was carried out to develop a task assignment algorithm for crowdsourcing software development that takes task types and personality types (e.g., Introversion and Extroversion, etc. ) into account. The proposed method functions in the same way as the open call format, but it incorporates personality-based categorization \cite{gilal2022task}.  To identify the personality types of software practitioners, the Myers-Briggs Type Indicator (MBTI) was used \cite{myers1985manual}. MBTI is a popular tool used to analyze personality types in the workplace, which measures a person across four dimensions:  energizing, attending, deciding, and living \cite{capretz2015influence}. Study S58 was carried out to find the relationship between testing technique, tester, and personality type and identify which personality types of testers are more competent at a particular testing approach \cite{kamangar2019enhance}. They also used the MBTI to identify the personality types \cite{capretz2015influence}.

Study S9 incorporated the assumption that \emph{sentiments} can influence newcomers in selecting issues to work on. Hence, they calculated the sentiments associated with the descriptions of the unprocessed issue. It returns two scores because people might express both positive and negative emotions in the same line, such as ``I loved this tool a lot until they turned it into worthless garbage of software''.
Study S59 posited that when faced with complex problems, novices who exhibit a more positive communication style are more likely to get assistance from expert developers.
Hence, they utilize the titles and descriptions of previous Pull Requests and issues to compute sentiment measurements for newcomers. The measurements incorporate the textual sentiment polarity mean and median values of their previously submitted PRs and issues that have been reported.

Demographic information, such as \emph{age, location, education, etc.}, are currently included in Studies S15 and S28, while Study S38 suggested considering the \emph{geographic region} in the future. Study S15 modeled developer and project profiles by analyzing their basic information and mining user online behavior logs. Then they filtered out the projects based on the location, age, and education level requirements. This filtration process was done to speed up the recommendation process. It has been observed that workers from certain countries tend to prefer collaborating with task posters from specific other countries. For example, workers from India often prefer to work with task posters from the US and the UK \cite{abhinav2020tasrec}. Hence, Study S28 captured a worker's preference for collaborating with task posters from certain countries. They extracted the worker's location, education, etc., from the worker's profile information.

The developer's \emph{discussion engagement level} was used in the recommendation system proposed in Study S12. It aimed to understand how developers contribute to communication. For example, some developers play a key role in discussions that may reflect their expertise or the importance of the topic, potentially correlating with specific API domains \cite{santos2023tell}. Study S1 suggested including the developer's level of engagement in discussions as a further improvement.

Study S26 considered \emph{engagement states (anxiety, apathy, and boredom)} when recommending questions. Study S10 suggested considering \emph{user satisfaction} as a future improvement to their recommendation system.
Study S29 considered software practitioner's \emph{motivation} when recommending tasks, while Studies S18 and S28 mentioned it as a future improvement. Studies S12, S28, and S29 suggested including the developer's \emph{career goals} as a future improvement to the proposed systems. Study S10 suggested including \emph{responsiveness}, which refers to how long it takes for the user to respond to the recommendation. The quicker the response is, the higher the responsiveness is.

\begin{center}
\begin{tcolorbox}[colback=gray!5!white,colframe=black!75!black]
\textbf{RQ3 Summary:} Human factors play a major role in software engineering. Hence, we analyzed the human factors currently being used in the existing recommendation systems and the human factors suggested to be included as future improvements to the systems platform (See table \ref{human factors}). Some of the human factors currently being used in the existing recommendation systems are personality types, sentiments, motivation, and career goals.
\end{tcolorbox}
\end{center}

\begin{table}[]
\caption{Human factors included and suggested to be included in the studies}
\label{human factors}
\centering
\begin{tabularx}{\textwidth}{XXX}
\toprule
\textbf{Human Factor} & \textbf{Studies covering human factor} & \textbf{Studies suggesting human factor} \\
\midrule
Topic and task type preference & S2, S13, S18, S31, S54, S55, S57, S58, S60 & S35 \\
\midrule
Skill preference & S1, S5, S15, S16, S32, S59, \textcolor{black}{S64} & N/A \\
\midrule
Social and behavioral preferences & S8, S10, S43, \textcolor{black}{S65} & S5 \\
\midrule
Personality & S4, S57, S58 & N/A \\
\midrule
Sentiments & S9, S59 & N/A \\
\midrule
Age, location, education level, gender & S15, S28 & S38, S49 \\
\midrule
Discussion engagement level & S12 & S1 \\
\midrule
Engagement status (Anxiety, Apathy, and Boredom) & S26 & N/A \\
\midrule
Motivation & S29 & S18, S28 \\
\midrule
Career goals & N/A & S12, S28, S29 \\
\midrule
User satisfaction & N/A & S10 \\
\midrule
Responsiveness & N/A & S10 \\
\bottomrule
\end{tabularx}
\end{table}

\subsection{RQ4: What platforms are currently being used to recommend crowdsourced software engineering tasks?}

Table \ref{platforms list} shows the crowdsourcing platforms used in the selected primary studies. 

\emph{GitHub} is the most popular platform used in the primary studies. It is used in \textcolor{black}{38} out of \textcolor{black}{65} studies (\textcolor{black}{58.4\%} of the studies). It is a developer-based platform to build, scale, and deliver secure software \cite{github}. Some of the Github projects used in the existing studies are JabRef, PowerToys, and Audacity. A further analysis of these projects is available in table \ref{github project scales}.

\emph{Topcoder} collaborates with clients to provide tailored solutions and support for both small tasks and large-scale projects throughout the entire software development process \cite{topcoder}. The Topcoder project categories used in the primary studies are BugBash, Code, UI, Prototype, and Quality Assurance. 

\emph{Bugzilla} is a comprehensive system for tracking defects or bugs in software. It helps developer teams manage and monitor issues, enhancements, and change requests in their products efficiently \cite{bugzilla}. Eclipse, Mozilla, and LibreOffice are the projects used from the Bugzilla platform in the primary studies.

\emph{Stack Overflow} is a website where programmers, both professionals and enthusiasts, ask and answer questions about programming. It aims to create a collection of thorough and top-notch answers for every programming-related question \cite{stackoverflow}.

\emph{JointForce}, created by ChinaSoft International, is a reliable IT service platform that uses social collaboration and a sharing economy model. It ranks 8th among China's top 100 software companies \cite{jointforce}.

\emph{uTest} is a platform where freelance software testers can learn, earn money, and contribute to the improvement of digital products for various brands \cite{utest_about}.

\emph{Amazon Mechanical Turk (MTurk)} is a crowdsourcing platform where people and businesses can hire remote workers to complete tasks and jobs online \cite{mturk}.

\emph{Testin} is a platform that offers testing, security, promotion, optimization, and AI solutions for applications. It serves over one million developers and enterprises globally \cite{testin}.

\emph{Baidu, Zhubajie, Oschina, Witkey, MoocTest, and Chinese open source crowdsourcing platform}, are popular Chinese crowdsourcing platforms used in the primary studies.

\begin{footnotesize}
\begin{longtable}{@{}p{0.3\textwidth}p{0.4\textwidth}p{0.2\textwidth}ll@{}}
\caption{Crowdsourcing platforms and projects}
\label{platforms list}\\
\toprule
\textbf{Platform} & \textbf{Projects} & \textbf{Studies} \\
\midrule
\endfirsthead
\multicolumn{5}{c}%
{{\tablename\ \thetable{} -- continued from previous page}} \\
\toprule
\textbf{Platform} & \textbf{Projects} & \textbf{Studies} \\
\midrule
\endhead
\midrule
\multicolumn{5}{r}{{Continued on next page}} \\
\endfoot
\bottomrule
\endlastfoot
\multirow{21}{=}{Github} & JabRef & S1, S12 \\
 & PowerToys & S1, S12 \\
 & Audacity & S1, S12 \\
 & Qt & S9 \\
 & Eclipse & S9, S11, S43 \\
 & LibreOffice & S9 \\
 & Mozilla & S11, S43 \\
 & CNCF projects & S16 \\
 & babel & S33 \\
 & Bitcoin & S33 \\
 & Jest & S33 \\
 & Lighthouse & S33 \\
 & Scaffold & S33 \\
 & Packer & S33 \\
 & Graphql-engine & S33 \\
 & Minikube & S33 \\
 & React-admin S33 & S33 \\
 & React-server S33 & S33 \\
 & Python scientific computing software eco system & S34 \\
 & GH Archive & S21 \\
 & Not specified & S5, S6, S7, S10, S13, S17, S18, S19, S20, S22, S23, S24, S30, S32, S35, S36, S47, S50, S51, S52, S53, S54, S55, S56, S59, S62, S63, \textcolor{black}{S64, S65} \\
\midrule
\multirow{6}{=}{Topcoder} & BugBash & S27 \\
 & Code & S27 \\
 & UI & S27 \\
 & Prototype & S27 \\
 & Quality Assurance & S27 \\
 & Not specified & S28, S29, S39 \\
\midrule
\multirow{4}{=}{Bugzilla} & Eclipse & S3 \\
 & Mozilla & S3 \\
 & LibreOffice & S3 \\
 & Not specified & S10, S37 \\
\midrule
Stackoverflow & Not applicable & S25, S26, S38, S49 \\
\midrule
Jointforce & Not specified & S8, S15, S61 \\
\midrule
Baidu CrowdTest crowdtesting platform & Not specified & S2, S31 \\
\midrule
Utest & Not specified & S31 \\
\midrule
Amazon Mechanical Turk (MTurk) & Not specified & S40 \\
\midrule
Testin & Not specified & S31 \\
\midrule
Zhubajie & Not specified & S14 \\
\midrule
Oschina & Not specified & S14 \\
\midrule
Witkey & Not specified & S14 \\
\midrule
MoocTest & Not specified & S31 \\
\midrule
Chinese open source crowdsourcing platform & H5 applications, WeChat applications & S46 \\
\end{longtable}
\end{footnotesize}

 Table \ref{github project scales} shows the scales of the projects available in GitHub used in the primary studies. The projects with less than 100 contributors were considered small-scale projects, projects with 100-500 contributors were considered medium-scale projects, and projects with more than 500 contributors were considered large-scale projects. There were 3 small-scale projects (15\%), 6 medium-scale projects (30\%), and 11 large-scale projects (55\%). We could not gather information about the project scales of the other platforms as they were not accessible. 

 \begin{center}
\begin{tcolorbox}[colback=gray!5!white,colframe=black!75!black]
\textbf{RQ4 Summary:} We observed that GitHub is the most popular \textcolor{black}{58.4\%} crowdsourced software engineering platform among the selected primary studies. Further, we observed that the majority (55\%) of the GitHub projects that were used to recommend tasks were large-scale projects with more than 500 contributors. Further, Bugzilla, Topcoder, and Utest are some other popular CSE platforms (see Table \ref{platforms list}).
\end{tcolorbox}
\end{center}

\begin{footnotesize}
\begin{longtable}{@{}p{0.2\textwidth}p{0.3\textwidth}p{0.35\textwidth}p{0.05\textwidth}ll@{}}
\caption{GitHub project scales}
\label{github project scales}\\
\hline
Github Project &
  Link &
  Contributors count &
  Scale \\ \hline
\endfirsthead
\endhead
Python scientific computing software ecosystem &
  https://github.com/python/cpython &
  \begin{tabular}[c]{@{}l@{}}2619\\ (for the repository with the highest number of stars)\end{tabular} &
  large \\ \hline
Jest &
  https://github.com/jestjs/jest &
  1525 &
  large \\ \hline
LibreOffice &
  https://github.com/LibreOffice/core &
  1251 &
  large \\ \hline
Packer &
  https://github.com/hashicorp/packer &
  1202 &
  large \\ \hline
babel &
  https://github.com/babel/babel &
  1071 &
  large \\ \hline
Bitcoin &
  https://github.com/bitcoin/bitcoin &
  938 &
  large \\ \hline
Minikube &
  https://github.com/kubernetes/minikube &
  861 &
  large \\ \hline
CNCF projects &
  https://github.com/cncf/landscape &
  \begin{tabular}[c]{@{}l@{}}794\\ (for the repository with the highest number of stars)\end{tabular} &
  large \\ \hline
JabRef &
  https://github.com/JabRef/jabref &
  614 &
  large \\ \hline
React-admin &
  https://github.com/marmelab/react-admin &
  580 &
  large \\ \hline
Graphql-engine &
  https://github.com/hasura/graphql-engine &
  546 &
  large \\ \hline
PowerToys &
  https://github.com/microsoft/PowerToys &
  433 &
  medium \\ \hline
Mozilla &
  https://github.com/mozilla/pdf.js &
  \begin{tabular}[c]{@{}l@{}}374\\ (for the repository with the highest number of stars)\end{tabular} &
  medium \\ \hline
Lighthouse &
  https://github.com/GoogleChrome/lighthouse &
  332 &
  medium \\ \hline
Qt &
  https://github.com/qt/qt &
  231 &
  medium \\ \hline
Audacity &
  https://github.com/audacity/audacity &
  203 &
  medium \\ \hline
Eclipse &
  https://github.com/eclipse/mosquitto &
  \begin{tabular}[c]{@{}l@{}}131\\ (for the repository with the highest number of stars)\end{tabular} &
  medium \\ \hline
Scaffold &
  https://github.com/scaffold-eth/scaffold-eth &
  94 &
  small \\ \hline
React-server &
  https://github.com/redfin/react-server &
  72 &
  small \\ \hline
GH Archive &
  https://github.com/igrigorik/gharchive.org &
  53 &
  small \\ \hline
\end{longtable}
\end{footnotesize}

\subsection{RQ5: Which features of platforms can be used to recommend crowdsourced software engineering tasks?}

\textcolor{black}{Features of the platform include structured elements such as issue or task descriptions, pull requests, comments, labels, etc.; these are tangible, system-generated artifacts within the development platform.}
Table \ref{attributes used to extract data} shows the features and the different attributes of the features of crowdsourcing platforms used to extract data. 

 \emph{Issue or task description} is the most popular feature used to extract different attributes. It was used in 30 primary studies. Title and description, attached files or links to other related issues, and assignee were extracted from the issue or task description. For example, S1 extracted the title and body from the issue descriptions. 

\emph{User activities} such as starring, forking, watching, and bidding were extracted in \textcolor{black}{24} primary studies. This is the second most popular feature used in primary studies. For example, Study S7 extracted starring and forking repositories in the GitHub platform to identify user preferences. 

\emph{Pull requests} are the third most popular feature used to extract different attributes. It was used in \textcolor{black}{18} primary studies. Commit message text, names of the files changed in the pull request, and pull request comments were extracted from the pull requests. For example, Study S1 extracted the name of the files changed, and commit messages from the pull requests to identify the relevant domain for the issues.

\emph{Comments} were used in \textcolor{black}{14} primary studies.  commenter's name and the text were extracted from the comments. Study S12 extracted commenter names to get the total number of commenters which they used as a social metric in their recommendation system.  

\emph{Labels and tags} provide insightful data about the tasks. Hence, they were used in 13 studies. Issue types such as verified, fixed, invalid, duplicate, wontfix, worksforme, critical, difficulty levels such as good first issue, medium difficulty, etc., tagged keywords were extracted from labels and tags. For example, Study S2 used bug labels and duplicate labels to identify the bug status.

\emph{Code} was used in  7 studies. APIs used in the code, variables, classes, identifiers, code snippets, and components or methods related to the tasks were extracted from the code. For example, Study S1 analyzed the code to identify the APIs used in the code as they recommend tasks based on App Programming Interface domains.

\emph{ReadME} file was used to extract data in 5 studies. The purpose and the usage of the project, information about different methods used in the project, and the copyright information were extracted from the Readme files of projects. For example, Study S17 extracted the purpose, usage, and programming language used in the project from the ReadMe file.

Further, \emph{Mailing lists, forums, source code management systems (SCM), and wikis} were also used to extract data in Study S41. 

\begin{center}
\begin{tcolorbox}[colback=gray!5!white,colframe=black!75!black]
\textbf{RQ5 Summary:} Issue or task description is the most popular feature used to extract data in primary studies. Further, pull requests, comments, labels, tags, etc., are other popular features used to extract data (see Table \ref{attributes used to extract data}). From those features, different attributes, such as commit message, task type, APIs used, etc, were extracted.
\end{tcolorbox}
\end{center}

\begin{footnotesize}
\begin{table}[htbp]
\caption{Platform features and attributes used to extract data}
\label{attributes used to extract data}
\centering
\begin{tabular}{p{0.3\textwidth} p{0.4\textwidth} p{0.2\textwidth}}
\toprule
Feature & Attributes & Studies \\
\midrule
Issue/task description & Title and description & S1, S3, S5, S7, S9, S11, S12, S14, S15, S27, S28, S30, S33, S34, S35, S37, S38, S39, S41, S42, S43, S44, S45, S46, S49, S54, S55, S59, S63 \\
 & Attached files or links & S3 \\
 & Assignee & S3, S48 \\
\midrule
User activities & Fork, star, watch, bid, etc & S5, S7, S16, S18, S19, S20, S21, S22, S23, S24, S32, S34, S35, S36, S44, S47, S50, S51, S56, S57, S62, S63, \textcolor{black}{S64,S65\%} \\
\midrule
Pull Requests & Commit messages & S1, S10, S11, S12, S18, S20, S33, S36, S41, S44, S53, S59, \textcolor{black}{S65} \\
 & Names of the files changed & S1, S12 \\
 & Pull request review comment & S2, S16, S21, S43, S51 \\
\midrule
Comments & Commenter name & S12 \\
 & Comment text & S1, S6, S10, S12, S13, S21, S33, S35, S36, S41, S43, S44, \textcolor{black}{S64, S65} \\
\midrule
Labels, tags & Issue or task type & S2, S3, S34, S37, S40, S41, S42, S43, S48, S51, S59 \\
 & Tagged keywords & S25, S61 \\
\midrule
Code & APIs used & S1 \\
 & Variables, classes, identifiers, code snippets & S13, S33, S34, S59 \\
 & Related components or methods & S17, S48 \\
\midrule
Readme file & Purpose and usage, methods information, copyright information & S17, S50, S52, S56, S59 \\
\midrule
Mailing lists, forums, source code management systems (SCM) and wikis. & N/A & S41 \\
\bottomrule
\end{tabular}
\end{table}

\end{footnotesize}

\section{Threats to Validity} \label{threats to validity}
Although we adhered closely to the recommendations outlined in \cite{kitchenham2007guidelines}, our SLR may have similar threats to validity as other SLRs in software engineering. The outcomes of our SLR might have been affected by the following validity threats.

\textcolor{black}{
\newline
\noindent \textbf{Internal validity.} To mitigate threats to internal validity, we adhered to a predefined SLR protocol. The search string was iteratively refined and tested before execution in order to optimize the results. Further, we developed our search string according to PICOC criteria, which enhances its comprehensiveness. To minimiz bias in the study selection process, we followed a multi-step filtering process. The first two authors validated the inclusion and exclusion criteria on a small subset of primary studies before applying them broadly. Papers were screened in several rounds: initially based on titles, abstracts, and keywords followed by a brief reading, and finally, a detailed review. Any disagreements regarding inclusion were resolved through discussions between the first and second authors. For data extraction, we developed a structured data extraction form (see Table \ref{table:8}) to ensure consistency. Since the first author conducted most of the extraction, regular discussions were held with the second author to clarify uncertainties and resolve discrepancies. A subset of the extracted data was further verified by the second and third authors.
\newline
\newline
\textbf{Construct validity.} To reduce threats to construct validity, we searched across six major digital libraries and employed two complementary search strategies: automatic database searches and snowballing. The selected primary studies provide highly relevant solutions to recommend a CSE task to a software practitioner. To further refine our selection process, we conducted several rounds of discussions to improve the inclusion and exclusion criteria.
\newline
\newline
\textbf{Conclusion validity.} To minimize biases in data synthesis and analysis, we adopted a collaborative approach. All authors participated in multiple rounds of discussions to determine the most effective way to categorize and present findings. This iterative process helped reduce subjective interpretations and ensured that our conclusions were well-grounded in the extracted data.
\newline
\newline
\textbf{External validity.} External validity reflects the generalizability of our findings. A potential limitation is the scope of the literature included. Our study relied on papers retrieved from six major digital libraries, including Scopus, which is known for its extensive indexing capabilities. However, studies outside these databases may not have been captured. To mitigate this, we employed snowballing techniques, manually reviewing references from selected papers to identify additional relevant studies. Further, we did not impose any time limitation on our search to capture all developments in the area. To improve the generalizability of our findings, future work could incorporate additional sources. 
}

\section{\textcolor{black}{Related Work}}
\label{Related Work}
During this review, we only considered peer-reviewed and published work and excluded blogs and media articles. Hence, we found one related existing systematic literature review and one survey study. We summarize the key aspects of these studies below.
\par Zhen et al. \cite{zhen2021crowdsourcing} conducted an SLR that focuses on the software tasks in the crowdsourcing domain at a high level, along with the tasks in non-software engineering domains such as data entry, writing, editing, etc. \textcolor{black}{The study examined the impact of task assignment in crowdsourcing, including the consequences of assigning a task to an appropriate worker and an inappropriate worker. The study highlights that improper task-worker matching not only affects result quality but also wastes valuable resources such as time, money, and client trust, underscoring the need for an optimal task assignment model. It emphasized the application of crowdsourcing in software engineering and analyzed existing methods for effective task allocation. Further, it identified 52 existing crowdsourcing platforms such as Amazon MechanicalTurk (AMT)\footnote{ \url{https://www.mturk.com/}}, Topcoder\footnote{ \url{https://www.topcoder.com}}, CastingWords\footnote{ \url{http://castingwords.com/}}.} Our SLR provides a comprehensive, in-depth analysis of personalized task recommendations to software practitioners in crowdsourcing platforms, focusing on recommending a task to a particular software practitioner in a crowdsourcing platform. Further, the existing SLR is conducted on the studies published from 2010 to 2019, whereas our SLR has no time duration restriction. Our search string significantly differs from the one used in \cite{zhen2021crowdsourcing}. They identified 120 relevant studies and selected 70 most relevant studies for their SLR. Only one primary study overlaps with the selected primary studies of our SLR. Regarding the research questions, our RQ1, RQ2, and RQ4 partially overlap with the existing literature review's  RQ3, RQ2, and RQ4, respectively. Moreover, we focus on assigning a suitable task to a particular software practitioner, whereas the majority of primary studies in the existing SLR focus on assigning a practitioner to a task. Since the existing SLR is a high-level study on the software engineering domain, the results include those unrelated to the software engineering field. Hence, there is a significant difference between our SLR and the existing SLR.
\par Mao et al. \cite{mao2017survey} performed a survey that focuses on how crowdsourcing is used in different stages of the software development life cycle along with its taxonomy. Further, it provides definitions for crowdsourced software engineering, crowdsourcing practices in software engineering, commercial platforms, and relevant case studies. \textcolor{black}{The study highlighted that while many previous studies have cited benefits such as low cost and quick delivery, there is a lack of comparative research that validates these advantages against traditional development methods. It also highlighted the significance of quality control in crowdsourced projects. Further, the findings of the study emphasize the challenges of managing communication, expectations, and workflows among crowd contributors with varying skill levels and motivations. It also highlights concerns about intellectual property rights and the need for clear guidelines to protect both requesters and contributors. Additionally, the study notes that cultural and regional differences can affect collaboration, making it crucial to understand these differences for effective global crowd management and alignment with project goals.} However, the survey is not based on any research questions as it is a broad analysis. They conducted the survey on the studies published from January 2006 to March 2015, whereas our SLR has no time duration restriction. They surveyed 210 publications out of 476 studies identified during the initial stage. There are no overlapping primary studies with our selected primary studies set for this SLR. Our search string significantly differs from the search string used in this survey. RQ4 in our SLR overlaps with the section ``4.1 commercial platforms'' in the survey. However, our SLR is focused on recommending crowdsourced software engineering tasks to software practitioners, whereas this survey is a broad analysis of crowdsourced software engineering. Hence, there is a significant difference in this survey with our SLR.

\section{Discussion and Future Research Directions} \label{discussion}
The research on task recommendation in Crowdsourced Software Engineering (CSE) is gaining increasing attention within the software engineering community. This is evidenced by a consistent upward trend in the volume of papers dedicated to this domain in the past decade (see Figure \ref{Publication trends}). Notably, over half of the papers reviewed (32 papers, 50\%) were published within the last three years. Several categories of software tasks have been recommended in several crowdsourcing platforms like GitHub and Topcoder using a large number of features and attributes in those platforms. It is crucial to systematically review and comprehensively document the various software task recommendation methods in crowdsourcing platforms to help understand the current progress of the area and potential future directions.  Based on the insights of this SLR, along with the identified limitations, we have pinpointed several key challenges in the domain of recommending tasks to software practitioners on crowdsourcing platforms. We outline these challenges below as recommendations to the Software Engineering research community for conducting further research on this domain.

\begin{itemize}
    \item \textbf{Human factors in software task recommendation systems}:
\par   
Software is developed by humans, read by humans, and maintained by humans \cite{bourque1998guide}. Human factors play a crucial role in software development \cite{curtis1988field}. Human factors affect software development team members, their activities, and ultimately, the quality of the product \cite{dutra2021human}. \textcolor{black}{Ockiya and Lock \cite{ockiya2023review} discovered a connection between human factors in remote software teams and their success or failure. Gaining a deeper understanding of this relationship and managing it effectively could help alleviate the challenges faced by remote teams. 
Some major challenges in CSE include ensuring effective communication and coordination, handling knowledge sharing, fostering motivation, building a network of volunteers, establishing trust, and facilitating team growth \cite{stol2014two, bhatti2020general, hosseini2014towards}.}
\par
Human factors have been integrated into some existing primary studies, and several studies plan to incorporate them as future enhancements to their recommendation systems. We analyzed the human factors currently being used or suggested to be used in the existing systems in Table \ref{human factors}. For example, Studies S9 and S59 used software practitioners' sentiment data in their recommendation system. Study S29 considers software practitioners' motivation when recommending tasks, whereas Studies S18 and S28 plan to consider motivation as a future improvement. Studies S12, S28, and S29 mentioned considering software practitioners' career aspirations in their recommendation systems as intended future improvements. Study S38 suggests including geographic regions, while Study S49 suggests demographic information.  Hence, this reflects that there is an interest among researchers to explore human factors affecting their recommendation systems. 
\par
Further, there are numerous ongoing studies on human factors in the software engineering domain. Researchers could significantly enhance their recommendation systems by incorporating these findings. For example, Dutra et al. identified 101 human factors that impact software development activities from various perspectives \cite{dutra2021human}. Further, Ortu et al. explored the correlation between the attractiveness of a project and the level of politeness exhibited by developers involved in its development \cite{ortu2015bullies}. They found that the more polite developers were, the more eager new developers were to collaborate on a project, and the longer they were willing to work on the project \cite{ortu2015jira}. Guzman et al. found that the commit comments written on Mondays tend to contain more negative emotions than positives \cite{guzman2013towards}. 
\par
Considering such human factors when recommending the tasks will improve the productivity of the practitioners \cite{guzman2013towards}. Additionally, promoting team diversity in terms of age, gender, disability, or ethnicity influences its positive impact on team member innovation \cite{dutra2021human}. Hence, these factors can be considered when onboarding newcomers to projects to maintain team diversity. Dutra et al. further mention that considering human factors helps in lowering the churn rate of software practitioners after onboarding to a project \cite{dutra2021human}.
\par
However, the main challenge of incorporating human factors into existing recommendation systems is the lack of data to extract human factors. Studies S4 and S57 tried, to some extent, to bridge this gap by introducing a new web platform to gather software practitioners' personality data. Further, gathering demographic information of software practitioners, such as age, gender, etc, may raise privacy and ethical concerns. For example, Study S4 mentioned that none of the vendors were willing to provide them with information regarding their clients' or developers' details. Privacy and ethical concerns could be the reason behind this situation. Hence, collecting such demographic information should be done with necessary careful handling. 
\par
It is clear that different human factors that are still undetermined could be used to recommend software tasks in crowdsourcing platforms. Hence, we encourage researchers to do an in-depth investigation on integrating human factors into the existing recommendation systems and the potential challenges that may stem from this integration.

\item\textbf{Transferring knowledge from one platform to another:}  
\par
\textcolor{black}{ As new software crowdsourcing platforms often lack sufficient historical data on developer behaviour, existing recommendation algorithms struggle to effectively match developers with new tasks \cite{yu2019cross}. Among the existing algorithms, collaborative filtering (CF) has gained significant recognition due to its versatility, interpretability, and independence from content-based features. However, it is also highly susceptible to data sparsity and the cold start problem. To address these challenges, CF algorithms with transfer learning have received considerable attention \cite{cantador2015cross, cremonesi2011cross}. By leveraging knowledge from data-rich auxiliary domains and transferring it to target domains with limited data, these systems help mitigate data sparsity issues and enhance recommendation accuracy. This transfer of knowledge enables new platforms to leverage existing user preferences, task information, and interaction patterns, thus enhancing their overall performance and user experience \cite{yan2017transfer} \cite{ying2017application} \cite{yan2013friend}. Given that contributors in CSE come from diverse backgrounds with varying levels of experience, leveraging prior data from different platforms can help match developers to projects more effectively. }

 For example, Studies S14, S15, and S27 mentioned addressing the  “cold start” problem by transferring knowledge or merging more information from other platforms as future improvements to their systems. Researchers in other domains have researched transferring knowledge from other platforms for their recommendation systems. For example, Yan et al. investigated cross-platform social relationships and behavior information to tackle the cold-start problem in friend recommendation \cite{6607510}. Specifically, they performed comprehensive data analysis to determine which information is most effectively transferable between platforms \cite{6607510}. Pan et al. addressed the data sparsity issue in the recommendation systems by transferring knowledge about both users and items from auxiliary data sources \cite{pan2010transfer}. Li et al. addressed the sparsity problem in individual rating matrices by learning a generative model from multiple related recommender systems \cite{li2009transfer}.
\par
Transferring knowledge can help initialize the recommendation engine and make informed recommendations even in the absence of historical data specific to the new platform  \cite{yan2017transfer}. Existing data from one platform can be used to bootstrap the recommendation system on another platform as it may reduce the need for collecting large amounts of new data \cite{yan2013friend}. Further, utilizing user preferences and interaction patterns from other platforms may allow for more personalized and relevant recommendations, enhancing user satisfaction \cite{zhang2021selective}. 
\par
However, sharing user data across platforms may raise privacy concerns \cite{jeckmans2013privacy}. Data breaches or unauthorized access during the transfer process can compromise sensitive information. Hence, careful consideration of data security measures and compliance with regulations is required. Further, it can be a technically complex task as it requires compatibility between different platforms, data formats, and systems \cite{yan2017transfer} \cite{min2015cross} \cite{fu2024exploring}.
\par
 It can be seen that only 4.7\% of primary studies have identified this area for improvement, and it is still largely unexplored. There is an opportunity for more studies to explore transferring knowledge from one platform to another. Hence, we recommend that researchers consider integrating knowledge transferring from other platforms into their recommendation systems.

\item \textbf{Cross-platform recommendation systems:} 
\par
Cross-platform recommendation systems provide users with personalized recommendations regardless of the platform they are signed up with.\textcolor{black}{These systems enable a broader reach and cater to the diverse needs of users who participate in crowdsourcing platforms.}
Only 4.6\% of the primary studies were tested on multiple crowdsourcing platforms. When the recommendation systems are designed to cater to a specific platform, those systems may be constrained by the features and capabilities of the platform they are built for \cite{suhas2021recommendation}\cite{chang2023cross}. It may limit their flexibility and adaptability. Further platform-specific recommendation systems may contribute to user lock-in \textcolor{black}{where users may feel constrained to stay on a single platform due to a lack of integration with other services or options. For example, Xu et al. \cite{xu_cross_domain}  proposed a cross-platform developer recommendation algorithm based on feature matching. Experimental results demonstrate that the proposed algorithm outperforms existing recommendation algorithms across various evaluation metrics.}

\par
The flexibility of platform-independent solutions enables the recommendation system to cater to diverse user needs and preferences without being tied to a specific platform's constraints. The interoperability of these solutions facilitates collaboration and data exchange between platforms, enabling more comprehensive and holistic recommendation strategies \cite{kanungo2020cross}. \textcolor{black}{This is particularly important in crowdsourcing, where users often contribute to projects across different platforms and bring a variety of skills, motivations, and preferences to the table.}

\par
However, developing cross-platform recommendation systems could be challenging as some required data may not be available on every platform in the same form \cite{chang2023cross} \cite{cao2017cross}. For example, the Topcoder platform offers money as a prize to the winners of their tasks, whereas GitHub does not provide such prizes for every task \cite{github} \cite{topcoder}. Hence, considering offered developer incentives as a recommendation factor in a cross-platform recommendation system is a complex task. On the other hand, when extending recommendation systems to work across multiple platforms, there's a risk of sacrificing the unique features of each platform, which could offer valuable insights for generating recommendations. For example, Studies S7 and S18 used the ``GitHub's forking a repository'' feature to extract user preferences. When extending those recommendation systems to cater to multiple platforms, they may not use such unique platform features to extract data.
\par
We encourage the researchers to build cross-platform recommendation systems to recommend software tasks to software practitioners in crowdsourcing platforms.

\item\textbf{Comprehensive evaluation, standardized evaluation metrics, and benchmarking are required in future studies: } 
\par
Evaluation of the existing systems often lack comprehensiveness due to the narrow focus on technical metrics, limited data sets, limited real-world studies, etc. Further evaluations covering more aspects of the proposed recommendation systems were suggested in the primary Studies S2, S11, S13, S18, S20, S23, S34, S37, S48, S50, and S62. Study S1 assesses the introduced tool's impact where they gathered feedback from contributors and analyzed how it influenced their choices through controlled experiments. Studies S11, S13, S18, S48, and S50 plan to broaden experiments to encompass more datasets, aiming to mitigate potential threats to external validity. Study S20 plans to conduct surveys involving junior and senior developers to ascertain the effectiveness of the introduced metrics, especially those related to commenting activities, and apply them to solve various problems. Study S23 intends to perform survey-based online experiments with actual developers to assess the usefulness of the proposed recommendation system. Study S34 plans to evaluate the recommendation model on larger projects in future research, including both open-source and commercial projects, to ensure a comprehensive evaluation. Study S37  plans to conduct user studies to evaluate the usability and effectiveness of Tesseract, particularly focusing on its search features. Study S62 aims to apply their proposed model to experiment on larger and richer datasets to validate its performance. 
\par
These further evaluations help to understand current gaps in the existing literature.
Establishing standardized evaluation metrics, datasets, and benchmarking frameworks for comparing and assessing the performance of different recommendation approaches could facilitate reproducibility, transparency, and rigor in evaluating the effectiveness of software task recommendation systems across various settings and domains \cite{sun2020we}. Further evaluations covering multiple aspects of recommendation systems, including technical metrics, real-world studies, and user feedback, provide more comprehensive recommendations  \cite{zangerle2022evaluating}. Evaluating recommendation models on larger and richer datasets helps validate their performance in real-world settings, ensuring that they can scale effectively and provide accurate recommendations across different contexts \cite{said2012recommender} \cite{zangerle2022evaluating}. 

\par
However, conducting comprehensive evaluations, including user studies, surveys, and experiments on larger datasets, can be resource-intensive in terms of time, effort, and cost \cite{zangerle2022evaluating}.

\par
We recommend researchers do comprehensive evaluations when developing recommendation systems, and community to establish standardized evaluation metrics and benchmarking.

\item\textbf{Inegrating recommendation systems with development workflow tools:}
\par
Professional software developers dedicate about one-third of their time to working within an integrated development environment (IDE), making it the most frequently used application throughout their workday \cite{sillitti2012understanding}. Hence, numerous studies have been conducted to enhance these IDEs for a seamless development experience. For example, Bacchelli et al. introduced an Eclipse plugin for integrating Stack Overflow community knowledge directly into the IDE \cite{bacchelli2012harnessing}. Luca Ponzanelli developed an Eclipse IDE plugin that automatically fetches, assesses, and recommends Stack Overflow discussions to developers once a specific confidence threshold is exceeded \cite{ponzanelli2014holistic}.
\par
Integrating recommendation systems with development workflow tools involves embedding recommendation functionalities directly into software development tools and environments commonly used by developers, such as IDEs, version control systems, project management tools, or communication platforms. For example, Study S1 suggests integrating the proposed recommendation system with Github actions or bots. Study S3 developed an open source plugin for some of JetBrains’ open source IDEs such as IntelliJ IDEA, CLion, etc. 
\par
Integrating the recommendation system with development workflow tools streamlines the task discovery and assignment process, enabling developers to access relevant tasks directly within their existing development environment without needing to switch between different tools or platforms. \textcolor{black}{Further, current adaptive IDE systems, such as the one proposed by Schmidmaier et al. \cite{schmidmaier2019real}, can track the task context, developer's expertise level, and interaction patterns directly from the IDE. In crowdsourced development, contributors are diverse and work remotely. Unlike traditional teams, where roles and expertise are known, crowdsourcing platforms lack direct visibility into a developer’s real-time interests and skill level. By leveraging real-time information from IDEs, recommendation systems can refine their recommendations to better align with the developer’s context. For instance, if the IDE detects that a developer is working on a machine learning project and has an advanced expertise level, the system can prioritize recommending complex ML-related crowdsourced tasks. }

\par
However, different development tools and environments have varied architectures and APIs that make the integration process complex \cite{di2020democratizing}. For example, integrating with Eclipse IDE might differ significantly from integrating with Visual Studio IDE \cite{zhao2013developing}. This can be addressed by utilizing a plugin-based architecture where the recommendation functionalities are provided as plugins or extensions for different IDEs and tools \cite{zhao2013developing} \cite{chatley2005predictable}. Further, maintaining compatibility with continuously evolving tools and the latest versions of the IDEs can be challenging. Another key challenge of integrating recommendation systems with development workflow tools is that recommendation systems can consume significant computational resources, potentially slowing down the development environment \cite{gasparic2017context}. 

\par
We suggest researchers integrate their recommendation systems with development workflow tools.

\item\textbf{The need for interactive recommendation systems:}
\par
Interactive recommendation systems enable users to actively engage with the recommendation process by providing feedback, preferences, and additional input \cite{liu2020diversified}. \textcolor{black}{In the context of CSE, where contributors vary in expertise, availability, and motivations, interactive recommendations can help match developers with tasks more effectively.} Tailoring recommendations to individual preferences increases the relevancy of the recommendations. Continuous feedback may help refine and improve the recommendation algorithms over time. Hence, such interaction enhances user satisfaction and improves the user experience by adapting to changing preferences over time \cite{wu2024survey,he2016interactive}. For example, Study S3 currently allows users (specifically Eclipse developers) to provide feedback through likes, dislikes, or snoozes (reminders) for issues. They utilize this feedback to hide certain issues in case of dislikes and to re-rank the recommendation list based on likes. In the future, they intend to leverage this data to automatically enhance and adjust the recommendation model. 
\par
Hau Xuan Pham \& Jason J. Jung developed an interactive recommendation system designed to assist users in rectifying their own ratings. Their approach involved determining whether user ratings align with their preferences, and subsequently correcting these ratings to enhance recommendations \cite{pham2013preference}. Mahmood et al. introduced a novel recommender system that autonomously learns an adaptive interaction strategy to aid users in achieving their interaction objectives \cite{mahmood2007learning}. Hariri et al. introduced an interactive recommender system that can identify and adjust to changes in context by analyzing the user's ongoing behavior \cite{hariri2014context}.
\par
Additionally, the development of Large Language Models (LLMs), like ChatGPT and GPT-4, has transformed the fields of Natural Language Processing (NLP) and Artificial Intelligence (AI). These models exhibit exceptional proficiency in fundamental tasks of language understanding and generation, alongside notable generalization abilities and reasoning skills \cite{fan2023recommender}. \textcolor{black}{Study S64 mentioned in the future works, integrating LLMs like GPT with their recommendation system can significantly improve recommendation accuracy by capturing complex linguistic patterns and semantics. LLMs could assist in dynamically refining task recommendations by processing natural language queries, extracting intent from developer interactions, and suggesting relevant tasks that align with evolving skills and interests.} Hence, LLMs can be employed to provide interactive recommendations as those excel at processing and understanding natural language input and leverage context from user queries and previous interactions to generate more relevant and personalized recommendations \cite{lin2023can}. \textcolor{black}{LLMs can help address the cold-start problem by leveraging limited data to generate personalized, context-aware recommendations. Their ability to process customizable prompts in chat-based systems enhances user engagement. For example, the LLM-based recommendation system in \cite{zhang2023recommendation} allows users to express their preferences and intents in natural language. This approach outperforms traditional methods relying solely on user-item interactions. Given their effectiveness in tackling data sparsity and improving efficiency, language models have become a promising approach in advancing recommendation systems in both academia and industry \cite{wu2024survey}.}

\par 
However, implementing interactive features can add complexity to the system and increase computational overhead \cite{mahmood2007learning}. Further, interactive recommendation systems require collecting vast amounts of user data, including preferences, feedback, and interaction history. The extent and granularity of this data can raise concerns about user privacy \cite{he2016interactive}.
Hence, we suggest developing interactive recommendation systems for CSE task recommendation in the future.

\end{itemize}

A summary of future research directions and the potential challenges of those improvements are available in table \ref{future directions}.

\begin{table}[]
\caption{Future research directions and potential challenges.}
\label{future directions}
\resizebox{\textwidth}{!}{%
\begin{tabular}{@{}ll@{}}
\toprule
\textbf{Future Research Direction} &
  \textbf{Potential Challenges} \\ \midrule
Integrating human factors into recommendation systems &
  \begin{tabular}[c]{@{}l@{}}Lack of data to extract human factors.\\ Privacy and ethical concerns.\end{tabular} \\ \midrule
Transferring knowledge from one platform to another &
  \begin{tabular}[c]{@{}l@{}}Privacy concerns.\\ Data breaches or unauthorized access during the transfer process.\\ Technical complexity.\end{tabular} \\ \midrule
Introducing cross-platform recommendation systems &
  May not be able to use unique platform features. \\ \midrule
Comprehensive evaluation, standardized evaluation metrics, and benchmarking are required in future studies &
  Can be resource-intensive in terms of time, effort, and cost. \\ \midrule
Integrating recommendation systems with development workflow tools &
  \begin{tabular}[c]{@{}l@{}}Technical complexity.\\ Maintaining compatibility with continuously evolving tools and the latest versions.\\ May consume significant computational resources.\end{tabular} \\ \midrule
Developing interactive recommendation systems &
  \begin{tabular}[c]{@{}l@{}}May add complexity to the system and increase computational overhead.\\ Privacy concerns.\end{tabular} \\ \bottomrule
\end{tabular}%
}
\end{table}

\section{Conclusion} \label{conclusion}
This SLR has provided a detailed analysis of software task recommendations to software practitioners in crowdsourcing platforms. Our analysis equips researchers with a comprehensive understanding of existing research in the domain, including the existing methods, benefits, and limitations of those methods, different task types recommended, use of human factors in recommendations, popular crowdsourcing platforms, and the features of those platforms used to extract data. Further, we highlighted the key areas for future research. For the SLR, We adhered to the systematic process outlined by Kitchenham et al \cite{kitchenham2007guidelines}. We selected  highly relevant primary studies for the SLR after a rigorous filtration process. Our main findings indicate that the number of studies on crowdsourced software engineering task recommendations significantly increased over the past couple of years, and we anticipate that this trend will continue. The identified future directions for further research in this domain include: (1) More studies are needed to explore the effect of different human factors such as the personality of software practitioners, motivation, and career goals on CSE; (2) limited attention has been paid to transferring knowledge from other related platforms, which can be beneficial in addressing the  “cold start” problem; (3) Investigation is needed to introduce generalizable platform-independent recommendation systems; (4) Need to conduct comprehensive further evaluation on existing systems and need establish standardized evaluation metrics and bench-marking; (5) Investigation needed to integrate the recommendation system with development workflow tools; (6) Can develop existing systems to interactive recommendation systems with the help of large language models which enable users to actively engage with the recommendation process.
   
\section{Acknowledgments}
This work was supported by the Deakin University Postgraduate Research Scholarship (DUPR-ROUND 0000018830). 

\appendix

\section{\textcolor{black}{Primary Studies}}

See Table \ref{selected primary studies}  \label{appendix:a}

\section{\textcolor{black}{Quality Assesment Scores}}
\textcolor{black}{See Table \ref{qa_scores} \label{appendix:b}}

\begin{footnotesize}
\begin{longtable}{@{}p{0.5cm}p{0.5cm}p{6cm}p{5cm}p{1cm}@{}}
\caption{Selected Primary Studies}
\label{selected primary studies}\\
\toprule
Ref &
  ID &
  Title &
  Venue &
  Year \\* \midrule
\endfirsthead
\endhead
\cite{vargovich2023givemelabeledissues} &
  S1 &
  GiveMeLabeledIssues: An Open Source Issue Recommendation System &
  International Conference on Mining Software Repositories (MSR) &
  2023 \\* \midrule
\cite{wang2021context} &
  S2 &
  Context-Aware Personalized Crowdtesting Task Recommendation &
  IEEE Transactions On Software Engineering &
  2022 \\* \midrule
\cite{samer2019towards} &
  S3 &
  Towards Issue Recommendation for Open Source Communities &
  International Conference on Web Intelligence &
  2019 \\* \midrule
\cite{tunio2017impact} &
  S4 &
  Impact of Personality on Task Selection in Crowdsourcing Software Development: A Sorting Approach &
  IEEE Access &
  2017 \\* \midrule
\cite{xu2023personalized} &
  S5 &
  Personalized Repository Recommendation Service for Developers with Multi-modal Features Learning &
  International Conference on Web Services (ICWS) &
  2023 \\* \midrule
\cite{golla2022project} &
  S6 &
  Project Recommendation for Open Source Communities &
  International Conference on Emerging Research in Electronics, Computer Science and Technology (ICERECT) &
  2022 \\* \midrule
\cite{xu2017repersp} &
  S7 &
  REPERSP: Recommending Personalized Software Projects on GitHub &
  International Conference on Software Maintenance and Evolution &
  2017 \\* \midrule
\cite{li2016task} &
  S8 &
  Task Recommendation with Developer Social Network in Software Crowdsourcing &
  Asia-Pacific Software Engineering Conference &
  2016 \\* \midrule
\cite{stanik2018simple} &
  S9 &
  A Simple NLP-Based Approach to Support Onboarding and Retention in Open Source Communities &
  International Conference on Software Maintenance and Evolution &
  2018 \\* \midrule
\cite{wang2018recommendation} &
  S10 &
  A Recommendation Method for Social Collaboration Tasks Based on Personal Social Preferences &
  IEEE Access &
  2018 \\* \midrule
\cite{yang2016combining} &
  S11 &
  Combining Word Embedding with Information Retrieval to Recommend Similar Bug Reports &
  International Symposium on Software Reliability Engineering &
  2016 \\* \midrule
\cite{santos2023tell} &
  S12 &
  Tell Me Who Are You Talking to and I Will Tell You What Issues Need Your Skills &
  International Conference on Mining Software Repositories (MSR) &
  2023 \\* \midrule
\cite{kim2018semantic} &
  S13 &
  Semantic Similarity-Based Contributable Task Identification for New Participating Developers &
  Journal of Information and Communication Convergence Engineering &
  2018 \\* \midrule
\cite{yang2017cold} &
  S14 &
  Cold-Start Developer Recommendation in Software Crowdsourcing: A Topic Sampling Approach &
  International Conference on Software Engineering and Knowledge Engineering &
  2017 \\* \midrule
\cite{qiao2018reinforcement} &
  S15 &
  A Reinforcement Learning Solution to Cold-Start Problem in Software Crowdsourcing Recommendations &
  International Conference on Progress in Informatics and Computing (PIC) &
  2018 \\* \midrule
\cite{lin2022open} &
  S16 &
  Open Source Software Supply Chain Recommendation Based on Heterogeneous Information Network &
  International Symposium on Benchmarking, Measuring and Optimization &
  2023 \\* \midrule
\cite{liao2023graph} &
  S17 &
  Graph Convolutional Network-Based Repository Recommendation System &
  Computer Modeling in Engineering and Sciences &
  2023 \\* \midrule
\cite{sayce2022recommendation} &
  S18 &
  Recommendation System for Open Source Projects for Minimizing Abandonment &
  The International FLAIRS Conference Proceedings &
  2022 \\* \midrule
\cite{zhu2021open} &
  S19 &
  Open-source project recommendation model &
  E3S Web of Conferences &
  2021 \\* \midrule
\cite{cseker2021new} &
  S20 &
  New developer metrics for open source software development challenges:  An empirical study of project recommendation systems &
  Applied Sciences &
  2021 \\* \midrule
\cite{su2021github} &
  S21 &
  A GitHub Project Recommendation Model Based on Self-Attention Sequence &
  International Conference on Big Data Engineering &
  2021 \\* \midrule
\cite{yang2021improving} &
  S22 &
  Improving Personalized Project Recommendation on GitHub Based on Deep Matrix Factorization &
  International Conference on Collaborative Computing: Networking, Applications and Worksharing &
  2021 \\* \midrule
\cite{sun2018personalized} &
  S23 &
  Personalized project recommendation on GitHub &
  Science China Information Sciences &
  2018 \\* \midrule
\cite{xu2017scalable} &
  S24 &
  Scalable relevant project recommendation on GitHub &
  Asia-Pacific Symposium on Internetware &
  2017 \\* \midrule
\cite{fukui2019suggesting} &
  S25 &
  Suggesting Questions that Match Each User's Expertise in Community Question and Answering Services &
  International Conference on Software Engineering, Artificial Intelligence, Networking and Parallel/Distributed Computing (SNPD) &
  2019 \\* \midrule
\cite{mogavi2019hrcr} &
  S26 &
  HRCR: Hidden Markov-Based Reinforcement to Reduce Churn in Question Answering Forums &
  Pacific Rim International Conference on Artificial Intelligences &
  2019 \\* \midrule
\cite{fu2021tdmatcher} &
  S27 &
  TDMatcher: A topic-based approach to task-developer matching with predictive intelligence for recommendation &
  Applied Soft Computing &
  2021 \\* \midrule
\cite{abhinav2020tasrec} &
  S28 &
  TasRec: A Framework for Task Recommendation in Crowdsourcing &
  International Conference on Global Software Engineering &
  2022 \\* \midrule
\cite{abhinav2023crowdassist} &
  S29 &
  CrowdAssist: A multidimensional decision support system for crowd workers &
  Journal of Software: Evolution and Process &
  2021 \\* \midrule
\cite{ford2022reboc} &
  S30 &
  ReBOC: Recommending Bespoke Open Source Software Projects to Contributors &
  Symposium on Visual Languages and Human-Centric Computing (VL/HCC) &
  2022 \\* \midrule
\cite{ge2022leveraging} &
  S31 &
  Leveraging Android Automated Testing to Assist Crowdsourced Testing &
  IEEE Transactions on Software Engineering &
  2023 \\* \midrule
\cite{jiang2017open} &
  S32 &
  Open Source Repository Recommendation in Social Coding &
  International ACM SIGIR Conference on Research and Development in Information Retrieval &
  2017 \\* \midrule
\cite{huang2021characterizing} &
  S33 &
  Characterizing and predicting good first issues &
  International Symposium on Empirical Software Engineering and Measurement (ESEM) &
  2021 \\* \midrule
\cite{ren2023effective} &
  S34 &
  Effective Recommendation of Cross-Project Correlated Issues based on Issue Metrics &
  Asia-Pacific Symposium on Internetware &
  2023 \\* \midrule
\cite{xiao2022recommending} &
  S35 &
  Recommending good first issues in GitHub OSS projects &
  International Conference on Software Engineering (ICSE) &
  2022 \\* \midrule
\cite{yang2016repolike} &
  S36 &
  RepoLike: personal repositories recommendation in social coding communities &
  Asia-Pacific Symposium on Internetware &
  2016 \\* \midrule
\cite{wang2011bug} &
  S37 &
  Which bug should I fix: helping new developers onboard a new project &
  International Workshop on Cooperative and Human Aspects of Software Engineering &
  2011 \\* \midrule
\cite{fu2020user} &
  S38 &
  User intimacy model for question recommendation in community question answering &
  Knowledge-Based Systems &
  2020 \\* \midrule
\cite{karim2018learn} &
  S39 &
  Learn or earn? Intelligent task recommendations for competitive crowdsourced software development &
  Hawaii International Conference on System Sciences &
  2018 \\* \midrule
\cite{yuen2015probabilistic} &
  S40 &
  Probabilistic matrix factorization with active learning for quality assurance in crowdsourcing systems &
  International Conference WWW/Internet &
  2015 \\* \midrule
\cite{leban2013semantic} &
  S41 &
  Semantic tools for improving software development in open source communities &
  International Semantic Web Conference &
  2013 \\* \midrule
\cite{dai2023graph} &
  S42 &
  Graph collaborative filtering-based bug triaging &
  The Journal of Systems and Software &
  2023 \\* \midrule
\cite{wu2022spatial} &
  S43 &
  A spatial–temporal graph neural network framework for automated software bug triaging &
  Knowledge-Based Systems &
  2022 \\* \midrule
\cite{he2022gfi} &
  S44 &
  GFI-Bot: Automated Good First Issue Recommendation on GitHub &
  ACM Joint European Software Engineering Conference and Symposium on the Foundations of Software Engineering &
  2022 \\* \midrule
\cite{kashiwa2019raptor} &
  S45 &
  RAPTOR: Release-Aware and Prioritized Bug-Fixing Task Assignment Optimization &
  International Conference on Software Maintenance and Evolution (ICSME) &
  2019 \\* \cmidrule(r){1-4}
\cite{yu2019software} &
  S46 &
  Software Crowdsourcing Task Allocation Algorithm Based on Dynamic Utility &
  IEEE Access &
  2019 \\* \midrule
\cite{zhang2014recommending} &
  S47 &
  Recommending Relevant Projects via User Behaviour: An Exploratory Study on Github &
  International Workshop on Crowd-based Software Development Methods and Technologies &
  2014 \\* \midrule
\cite{yadav2022developer} &
  S48 &
  Developer load balancing bug triage: Developed load balance &
  Expert Systems &
  2022 \\* \midrule
\cite{fu2019tracking} &
  S49 &
  Tracking user-role evolution via topic modeling in community question answering &
  Information Processing and Management &
  2019 \\* \midrule
\cite{zhang2017detecting} &
  S50 &
  Detecting Similar Repositories on GitHub &
  International Conference on Software Analysis, Evolution and Reengineering (SANER) &
  2017 \\* \midrule
\cite{guendouz2015recommending} &
  S51 &
  Recommending Relevant Open Source Projects on GitHub using a Collaborative-Filtering Technique &
  International Journal of Open Source Software and Processes &
  2015 \\* \midrule
\cite{koskela2018open} &
  S52 &
  open source software recommendations using github &
  International Conference on Theory and Practice of Digital Libraries &
  2018 \\* \midrule
\cite{liu2018recommending} &
  S53 &
  Recommending GitHub Projects for Developer Onboarding &
  IEEE Access &
  2018 \\* \midrule
\cite{zhou2021ghtrec} &
  S54 &
  Ghtrec: A personalized service to recommend github trending repositories for developers &
  International Conference on Web Services (ICWS) &
  2021 \\* \midrule
\cite{kim2021sequential} &
  S55 &
  Sequential recommendations on github repository &
  Applied Sciences &
  2021 \\* \midrule
\cite{nguyen2018crosssim} &
  S56 &
  CrossSim: exploiting mutual relationships to detect similar OSS projects &
  Euromicro Conference on Software Engineering and Advanced Applications &
  2018 \\* \midrule
\cite{gilal2022task} &
  S57 &
  Task Assignment and Personality: Crowdsourcing Software Development &
  Research Anthology on Agile Software, Software Development &
  2022 \\* \midrule
\cite{kamangar2019enhance} &
  S58 &
  To enhance effectiveness of crowdsource software testing by applying personality types &
  International Conference on Software and Information Engineering &
  2019 \\* \midrule
\cite{xiao2023personalized} &
  S59 &
  Personalized First Issue Recommender for Newcomers in Open Source Projects &
  International Conference on Automated Software Engineering (ASE) &
  2023 \\* \midrule
\cite{yang2022crowdsourced} &
  S60 &
  Crowdsourced Testing Task Assignment based on Knowledge Graphs &
  International Conference on Software Quality, Reliability and Security (QRS) &
  2022 \\* \midrule
\cite{yan2017transfer} &
  S61 &
  Transfer learning for cross-platform software crowdsourcing recommendation &
  Asia-Pacific Software Engineering Conference (APSEC) &
  2017 \\* \midrule
\cite{fu2023graph} &
  S62 &
  Graph contextualized self-attention network for software service sequential recommendation &
  Future Generation Computer Systems &
  2023 \\* \midrule
\cite{abbasi2021relevant} &
  S63 &
  Relevant Projectrecommendation System Using Developer Behavior And Project Features &
  International Journal of Advanced Trends in Computer Science and Engineering &
  2021 \\* \midrule
  \cite{phatangare2024codecompass} &
  \textcolor{black}{S64} &
  \textcolor{black}{CodeCompass: NLP-Driven Navigation to Optimal Repositories} &
  \textcolor{black}{International Conference on Pervasive Computing and Social Networking} &
  \textcolor{black}{2024} \\* \midrule
  \cite{shen2024multi} &
  \textcolor{black}{S65} &
  \textcolor{black}{Multi-objective optimization and integrated indicator-driven two-stage project recommendation in time-dependent software ecosystem} &
  \textcolor{black}{Information and Software Technology} &
  \textcolor{black}{2024} \\* \bottomrule
\end{longtable}
\end{footnotesize}

\begin{table}[]
\caption{\textcolor{black}{Quality assessment scores for primary studies}}
\label{qa_scores}
\small
\begin{tabular}{@{}llllllllllll@{}}
\toprule
ID  & QA1 & QA2 & QA3 & QA4 & QA5 & ID  & QA1 & QA2 & QA3 & QA4 & QA5 \\ \midrule
S1  & 5   & 5   & 4   & 5   & 2   & S34 & 4   & 3   & 3   & 3   & 3   \\
S2  & 5   & 4   & 4   & 5   & 4   & S35 & 4   & 3   & 2   & 4   & 2   \\
S3  & 4   & 5   & 3   & 3   & 3   & S36 & 3   & 3   & 2   & 3   & 2   \\
S4  & 3   & 2   & 2   & 3   & 1   & S37 & 3   & 3   & 3   & 1   & 1   \\
S5  & 3   & 3   & 4   & 4   & 2   & S38 & 3   & 3   & 3   & 3   & 3   \\
S6  & 3   & 2   & 3   & 4   & 2   & S39 & 4   & 4   & 3   & 3   & 3   \\
S7  & 5   & 3   & 2   & 3   & 1   & S40 & 3   & 2   & 4   & 3   & 2   \\
S8  & 4   & 2   & 4   & 4   & 2   & S41 & 3   & 1   & 2   & 1   & 1   \\
S9  & 4   & 3   & 4   & 4   & 4   & S42 & 4   & 3   & 3   & 3   & 4   \\
S10 & 3   & 2   & 3   & 3   & 2   & S43 & 4   & 2   & 4   & 4   & 3   \\
S11 & 3   & 4   & 3   & 4   & 4   & S44 & 3   & 4   & 3   & 3   & 2   \\
S12 & 4   & 3   & 3   & 4   & 4   & S45 & 4   & 2   & 3   & 1   & 1   \\
S13 & 3   & 2   & 3   & 3   & 3   & S46 & 3   & 3   & 4   & 4   & 2   \\
S14 & 3   & 3   & 4   & 4   & 2   & S47 & 4   & 5   & 2   & 3   & 1   \\
S15 & 4   & 4   & 4   & 3   & 3   & S48 & 4   & 4   & 3   & 4   & 3   \\
S16 & 3   & 2   & 3   & 3   & 2   & S49 & 4   & 3   & 3   & 3   & 1   \\
S17 & 3   & 3   & 3   & 4   & 3   & S50 & 3   & 3   & 4   & 4   & 4   \\
S18 & 3   & 3   & 2   & 3   & 1   & S51 & 4   & 3   & 3   & 3   & 2   \\
S19 & 2   & 3   & 1   & 3   & 1   & S52 & 2   & 4   & 3   & 1   & 1   \\
S20 & 4   & 4   & 3   & 3   & 3   & S53 & 4   & 3   & 4   & 4   & 5   \\
S21 & 4   & 3   & 2   & 3   & 2   & S54 & 3   & 4   & 3   & 4   & 2   \\
S22 & 3   & 4   & 3   & 4   & 4   & S55 & 3   & 4   & 3   & 3   & 2   \\
S23 & 4   & 3   & 3   & 4   & 4   & S56 & 4   & 3   & 3   & 3   & 3   \\
S24 & 4   & 3   & 3   & 4   & 3   & S57 & 3   & 2   & 3   & 3   & 1   \\
S25 & 3   & 3   & 3   & 3   & 3   & S58 & 2   & 2   & 2   & 3   & 1   \\
S26 & 2   & 3   & 3   & 1   & 1   & S59 & 4   & 3   & 3   & 4   & 3   \\
S27 & 4   & 3   & 4   & 4   & 2   & S60 & 3   & 3   & 3   & 3   & 1   \\
S28 & 3   & 3   & 5   & 3   &     & S61 & 4   & 3   & 4   & 3   & 1   \\
S29 & 3   & 3   & 4   & 4   & 2   & S62 & 3   & 3   & 3   & 4   & 1   \\
S30 & 3   & 3   & 2   & 2   & 2   & S63 & 3   & 3   & 2   & 2   & 1   \\
S31 & 3   & 3   & 4   & 4   & 4   & S64 & 3   & 3   & 3   & 3   & 2   \\
S32 & 3   & 4   & 2   & 3   & 1   & S65 & 3   & 3   & 4   & 4   & 1   \\
S33 & 4   & 4   & 3   & 3   & 3   &     &     &     &     &     &     \\ \bottomrule
\end{tabular}%
\end{table}

\printcredits

\bibliographystyle{unsrtnat}

\bibliography{cas-refs}

\end{document}